%% file: main.tex
\documentclass{IEEEtaes}
\usepackage{cite}
\def\BibTeX{{\rm B\kern-.05em{\sc i\kern-.025em b}\kern-.08em
    T\kern-.1667em\lower.7ex\hbox{E}\kern-.125emX}}
\renewcommand{\appendixname}{APPENDIX}
\usepackage{amsmath}   
\usepackage{bbm}
\usepackage{cleveref}
\usepackage{adjustbox}

\usepackage{amsmath,amssymb,amsfonts}
\usepackage{algorithmic}
\usepackage{graphicx}
\usepackage{textcomp}
\usepackage{setspace} 
\usepackage{subcaption}
\usepackage{caption}
\usepackage{xcolor}
\usepackage[acronym]{glossaries}
\glsdisablehyper
\usepackage{booktabs,threeparttable}
\usepackage{array}
\usepackage{multirow}
\usepackage{multicol}
\usepackage[normalem]{ulem}
\input{acronyms}
\usepackage{color,array,amsthm}
\usepackage{graphicx}
\let\labelindent\relax
\usepackage{enumitem}
\usepackage{amsmath}

\usepackage{siunitx}
\usepackage{tabularx}
\newcommand{\fcl}{\bar{F1}_{\text{CL}}}
\newcommand{\pmv}[2]{$#1\text{\tiny$\pm#2$}$}
\newcommand{\pmvbf}[2]{$\text{\boldmath$#1$}\text{\tiny$\pm#2$}$}
\newcommand{\pmvul}[2]{$\underline{#1}\text{\tiny$\pm#2$}$}
\newcommand{\finedashuline}{\bgroup\setlength{\ULdepth}{1.2pt}\markoverwith{\kern.07em\vtop{\kern\ULdepth\hrule width .16em height 0.4pt}\kern.07em}\ULon}
\newcommand{\pmvdul}[2]{$\text{\finedashuline{$#1$}}\text{\tiny$\pm#2$}$}

\jvol{XX}
\jnum{XX}
\jmonth{XXXXX}
\paper{1234567}
\pubyear{2020}
\doiinfo{TAES.2020.Doi Number}

\begin{document}

\title{Overcoming Catastrophic Forgetting: Continual Learning for Radar Waveform Classification}

\author{Mohamed Rabie}

\author{Chinthana J. Panagamuwa}
\member{Member, IEEE}

\author{Konstantinos G. Kyriakopoulos}
\member{Senior Member, IEEE}
\affil{Wolfson School of Mechanical, Electrical and Manufacturing Engineering. Loughborough University, U.K.}

\receiveddate{This work has been submitted to the IEEE for possible publication. Copyright may be transferred without notice, after which this version may no longer be accessible.\\ \\
This study was supported by the EME Hub, funded by Defence Science and Technology Laboratory (Dstl).}


\authoraddress{Authors’ address: Mohamed Rabie, Chinthana Panagamuwa and Konstantinos G. Kyriakopoulos are with the Wolfson School of Mechanical, Electrical and Manufacturing Engineering at Loughborough University, UK. \itshape (Corresponding author: Mohamed Rabie \href{mailto:m.rabie@lboro.ac.uk}{(m.rabie@lboro.ac.uk)}).}


\markboth{RABIE ET AL.}{Overcoming Catastrophic Forgetting: Continual Learning for Radar Waveform Classification}
\maketitle

\input{Abstract.tex}


\input{01_Introduction/Introduction}
\input{02_Related_Work/Related_Work}

\input{03_Preliminaries/Preliminaries}

\input{04_Methodology/Methodology}
\input{05_Results/Results}
\input{Conclusion.tex}
\section*{ACKNOWLEDGEMENTS}
The authors acknowledge the use of Claude Opus for text editing, code implementation and debugging. Particularly, it was used to condense and clarify text, and for ensuring fairness across implementations. 
\input{07_Appendix/Appendix}

\bibliography{references_shortened}





\end{document}

%% file: acronyms.tex
\newacronym{mfr}{MFR}{Multi-Functional Radar}
\newacronym{ai}{AI}{Artificial Intelligence}
\newacronym{sjr}{SJR}{Signal-to-Jamming Ratio}
\newacronym{rl}{RL}{Reinforcement Learning}
\newacronym{drl}{DRL}{Deep Reinforcement Learning}
\newacronym{eme}{EME}{Electromagnetic Environment}
\newacronym{ecm}{ECM}{Electromagnetic Countermeasures}
\newacronym{rf}{RF}{Radio Frequency}
\newacronym{ew}{EW}{Electromagnetic Warfare}
\newacronym{cew}{CEW}{Cognitive Electromagnetic Warfare}
\newacronym{es}{ES}{Electromagnetic Surveillance}
\newacronym{mab}{MAB}{Multi-Armed Bandits}
\newacronym{sac}{SAC}{Soft-Actor Critic}
\newacronym{ddqn}{DDQN}{Double-Depth Q-Network}
\newacronym{dqn}{DQN}{Deep Q-Network}
\newacronym{ql}{QL}{Q-Learning}
\newacronym{a2c}{A2C}{Advantage-Actor Critic}
\newacronym{marl}{MARL}{Multi-Agent RL}
\newacronym{ppo}{PPO}{Proximal Policy Optimisation}
\newacronym{pri}{PRI}{Pulse Repetition Interval}
\newacronym{uav}{UAV}{Unmanned Aerial Vehicle}
\newacronym{per-ddqn}{PER-DDQN}{Prioritised Experience Replay DDQN}
\newacronym{dstl}{DSTL}{Defence Science and Technology Laboratory}
\newacronym{elint}{ELINT}{Electronic Intelligence}
\newacronym{tx-rx}{Tx-Rx}{transmitter-receiver}
\newacronym{bw}{BW}{Bandwidth}
\newacronym{pw}{PW}{Pulse Width}
\newacronym{lps}{LPS}{Learn-Prune-Share}
\newacronym{admm}{ADMM}{Alternating Direction Method of Multipliers}
\newacronym{rn}{ResNet}{Residual Network}
\newacronym{relu}{ReLU}{Rectified Linear Unit}
\newacronym{usrp}{USRPs}{Universal Software Radio Peripherals}
\newacronym{nist}{NIST}{National Institute of Standards and Technology}
\newacronym{awgn}{AWGN}{Additive White Gaussian Noise}
\newacronym{snr}{SNR}{Signal-to-Noise Ratio}
\newacronym{ds}{DS}{Dempster-Shafer}
\newacronym{lora}{LoRa}{Long Range Radio}
\newacronym{ll}{LL}{Lifelong Learning}
\newacronym{cnn}{CNN}{Convolutional Neural Network}
\newacronym{drc}{DRC}{Drone Remote Controllers}
\newacronym{rx}{Rx}{receiver}
\newacronym{rff}{RFF}{RF Fingerprinting}
\newacronym{iot}{IoT}{Internet of Things}
\newacronym{ml}{ML}{Machine Learning}
\newacronym{iq}{IQ}{In-phase and Quadrature}
\newacronym{mtl}{MTL}{Multi-Task Learning}
\newacronym{cl}{CL}{Continual Learning}
\newacronym{il}{IL}{Incremental Learning}
\newacronym{tid}{TID}{Task Identity}
\newacronym{sota}{SOTA}{state-of-the-art}
\newacronym{lpf}{LPF}{Low-Pass Filter}
\newacronym{comms}{Comms.}{communications}
\newacronym{st}{ST}{Single Task}
\newacronym{elc}{ELC}{Evidential Lifelong Classifier}
\newacronym{nn}{NN}{Neural Network}
\newacronym{ec}{EC}{Evidential Classifier}
\newacronym{blc}{BLC}{Bayesian Lifelong Classifier}
\newacronym{uq}{UQ}{Uncertainty Quantification}
\newacronym{ppb}{PPB}{Pulses per Burst}

\newacronym{fm}{FM}{Frequency Modulation}
\newacronym{lfm}{LFM}{Linear Frequency Modulation}
\newacronym{fsk}{FSK}{Frequency-Shift Keying}
\newacronym{psk}{PSK}{Phase-Shift Keying}
\newacronym{pm}{PM}{Phase Modulation}
\newacronym{cw}{CW}{Continuous Wave}
\newacronym{fmcw}{FMCW}{Frequency Modulated CW}
\newacronym{hfm}{HFM}{Hyperbolic FM}
\newacronym{nlfm}{NLFM}{Non-Linear FM}
\newacronym{qfm}{QFM}{Quadratic FM}
\newacronym{amc}{AMC}{Automatic Modulation Classification}

\newacronym{rml}{RML}{RadarML}
\newacronym{rc}{RC}{RadChar}
\newacronym{rnist}{RN}{RadNIST}
\newacronym{dr}{DR}{DeepRadar}

\newacronym{owm}{OWM}{Orthogonal Weight Modification}
\newacronym{aop}{AOP}{Adaptive Orthogonal Projection}
\newacronym{gem}{GEM}{Gradient Episodic Memory}
\newacronym{agem}{A-GEM}{Averaged GEM}
\newacronym{oml}{OML}{Online-aware Meta-Learning}
\newacronym{rln}{RLN}{Representation Learning Network}
\newacronym{pln}{PLN}{Prediction Learning Network}
\newacronym{nmn}{NMN}{Neuromodulatory Network}
\newacronym{pn}{PN}{Prediction Network}
\newacronym{fim}{FIM}{Fisher Information Matrix}
\newacronym{sl}{SL}{Supervised Learning}
\newacronym{ul}{UL}{Unsupervised Learning}
\newacronym{ssl}{SSL}{Semi-Supervised Learning}

\newacronym{bwt}{BWT}{Backward Transfer}
\newacronym{fwt}{FWT}{Forward Transfer}
\newacronym{ota}{OTA}{over-the-air}
\newacronym{sgd}{SGD}{Stochastic Gradient Descent}
\newacronym{lr}{LR}{Learning Rate}
\newacronym{amp}{AMP}{Automatic Mixed Precision}

\newacronym{cf}{CF}{Catastrophic Forgetting}
\newacronym{film}{FiLM}{Feature-wise Linear Modulation}
\newacronym{kl}{KL}{Kullback-Leibler}
\newacronym{til}{TIL}{Task-incremental Learning}
\newacronym{cil}{CIL}{Class-incremental Learning}
\newacronym{mlp}{MLP}{Multi-layer Perceptron}
\newacronym{qp}{QP}{Quadratic Program}
\newacronym{er}{ER}{Experience Replay}
\newacronym{er-res}{Res-ER}{Reservoir-sampling ER}
\newacronym{er-ring}{Ring-ER}{Ring-buffer ER}
\newacronym{ema}{EMA}{exponential moving average}
\newacronym{bn}{BN}{Batch Normalisation}
\newacronym{ci}{CI}{Confidence Interval}
\newacronym{hpt}{HPT}{Hyperparameter Tuning}
\newacronym{sesoi}{SESOI}{Smallest Effect Size of Interest}
\newacronym{ss}{SS}{statistically significant}
\newacronym{ns}{NS}{not significant}
\newacronym{ms}{MS}{marginally significant}
\newacronym{inc}{INC}{inconclusive}
\newacronym{ft}{FT}{Fine-Tuning}
\newacronym{jt}{JT}{Joint-Training}\newacronym{1e}{1E}{single-epoch}
\newacronym{5e}{5E}{multi-epoch}

%% file: Abstract.tex
\begin{abstract}
Automated radar waveform recognition classifies the waveform and modulation type of a received signal. As waveforms variants emerge appear over time, a deployed classifier must learn them without retraining from scratch. Catastrophic Forgetting occurs when learning additional waveforms overwrites prior knowledge. Continual Learning (CL) methods tackle this challenge, but prior radar studies only evaluate a few methods on narrow benchmarks. Practitioners lack evidence on which method to deploy or which mechanisms are effective. This work presents the first large-scale benchmark of continual learning for radar waveform recognition. Eighteen CL methods are evaluated on a sequential ten-task problem, drawn from four public datasets spanning 54 waveforms, under task- and class-incremental (TIL, CIL) settings and single- and multi-epoch regimes. Methods that retain data from old tasks lead both TIL and CIL settings. Plain replay alone closes 77\%--88\% of the gap between the lower and upper practical limits under TIL, but only 46\%--55\% under CIL. The internal mechanisms of the leading methods are ablated and statistically tested to identify effective mechanisms. Each ablation is paired against its full method over n=6 training seeds, topped up to n=12 when inconclusive, and p-values are Holm-Bonferroni corrected within each method's family. Ablation results show that several mechanisms confer smaller but statistically significant gains. Combining these mechanisms improves the leading F1 score by 3.4\%--5.5\% and 2.7\%--4.9\% under TIL and CIL, respectively, and the combined models exceed every benchmarked method in both settings for all SNRs. These results give researchers of radar waveform classification systems a mechanism-level basis for design.
\end{abstract}

%% file: 01_Introduction/Introduction.tex
\section{INTRODUCTION}
As the electromagnetic (EM) spectrum grows more congested, radar and communications systems increasingly employ parameter-agile and adaptive waveforms, widening the space a recogniser must handle. Deep learning approaches, particularly \glspl{cnn}, perform strongly on this task, learning discriminative features directly from \gls{iq} samples without expert-reliant feature engineering~\cite{2018_Kong_RWR-CNN}.

These models are trained on a fixed set of waveform classes. However, operational environments are not fixed---new and unforeseen signals emerge over time. The operator then faces three options, none of them satisfactory: i) Retraining on all accumulated data is computationally prohibitive at scale and assumes availability of old data, ii) maintaining an ensemble of specialist models adds management overhead and requires a priori knowledge for model selection, and iii) fine-tuning on new data alone is cheap but overwrites prior knowledge. This failure is known as \textit{\gls{cf}}~\cite{cf_1989}.

\Gls{cl}, also known as lifelong or incremental learning, addresses this failure directly. A \gls{cl} model learns new waveform classes incrementally whilst retaining prior ones, without retraining from scratch. Two settings are considered in this work. In the \gls{til} setting, waveform groups arrive as separate tasks with a classifier head each, and the \gls{tid} must be known at inference. In the \gls{cil} setting, new waveforms are added to a shared head, and the \gls{tid} is required during training but not at inference~\cite{cl_review2023}.

The \gls{cl} literature offers a broad family of solutions, spanning regularisation~\cite{ll_lwf_2017,ll_ewc_2017,2017_si,2018_rwalk}, replay~\cite{ll_exp_replay_2019,ll_gem_2017,ll_mof_2017}, optimisation~\cite{ll_gem_2017,AGEM,ll_lamaml_2020} and architectural isolation~\cite{ll_packnet_2018,2018_hat} approaches. However, their application to radar waveforms remains sparse. Firstly, existing studies evaluate a handful of methods, typically on fewer than ten classes drawn from a single dataset, and in a single incremental setting~\cite{xu2016_radar_il,2023_il-prototype_rei,2024_il-amc,2026_Rabie_ELC}. The relative merits of the method families are therefore unknown in this domain, leaving a practitioner without an evidence base from which to choose. Secondly, and more fundamentally, no study isolates \emph{which mechanism} within a method delivers its benefit. The leading methods each bundle several mechanisms, such as replay, gradient projection, meta-learning and distillation. Whole-method comparisons do not reveal which mechanisms are effective.

This work closes both gaps. Eighteen \gls{cl} methods are evaluated on four publicly available radar waveform datasets: RadarML~\cite{2025_White_RadarML}, RadNIST~\cite{ds_radnist}, RadChar~\cite{rc_Huang2023} and DeepRadar~\cite{ds_deeprad}, collectively spanning \num{54} unique waveforms, filtered to a common $-10$~dB to $+10$~dB \gls{snr} range. The datasets differ in modulation family, acquisition method, and sampling rate. A further constraint is that intercepts arrive as a continuous stream. At high sampling rates, raw \gls{iq} data cannot all be stored and revisited, so each sample may be trained on only once, as it arrives. Every method is therefore run under two regimes according to the number of passes permitted over each task's data: \gls{5e}, with five epochs per task, and \gls{1e}, a single pass. Comparing the two isolates how far each stability--plasticity trade-off degrades when data cannot be revisited. The contributions are as follows:
\begin{enumerate}[topsep=0pt]
    \item The first large-scale benchmark of \gls{cl} for radar waveform recognition, covering eighteen methods on \num{54} unique signals across ten tasks. In contrast, prior work typically uses ten or fewer classes in total~\cite{xu2016_radar_il,2023_il-prototype_rei,2024_il-amc,2026_Rabie_ELC}. Performance is assessed by the \gls{cl} F1-score $\fcl$, and backward and forward transfer.
    
    \item An evaluation of methods under both \gls{til} and \gls{cil} settings, as in \cite{vandeven2022three}, and under both \gls{1e} and \gls{5e}. The \gls{1e} regime evaluates how well new waveforms are learned from a stream in which each sample is seen once, as when intercepts cannot be stored and revisited. Leading methods are further evaluated at ${-}10~\text{dB}$ to $10~\text{dB}$ \gls{snr}.
    
    \item An attribution of performance to individual mechanisms rather than to whole methods, based on leave-one-out ablations of the leading methods with paired statistical significance testing. Each ablation removes or replaces a single mechanism, tested on $n={6}$ seeds and topped up to $n{=}12$ for inconclusive verdicts. Effects are Holm--Bonferroni corrected within each method and judged against a smallest effect size of interest of $1\%~\fcl$, so that a negligible effect can be told apart from an inconclusive one.
    
    \item Combined models that graft the significant mechanisms onto the leading methods. The mechanisms prove additive: Under \gls{til}, GEM$^+$ raises the leading $\fcl$ from $62.8\%$ to $66.2\%$ in \gls{1e}, and from $68.7\%$ to $74.2\%$ in \gls{5e}. Under \gls{cil}, BCL$^+$ raises the leading $\fcl$ by $2.7\%$--$4.9\%$. The memory and runtime costs of every method are also reported.
\end{enumerate}

Section~II positions the work within the radar waveform classification literature. Section~III introduces five families of \gls{cl} methods and details the $18$ methods in terms of their constituent mechanisms. Section~IV formalises the \gls{cl} problem, evaluation metrics and task construction, and sets out the ablations, significance testing and training setup. Section~\ref{sec:results} reports the benchmark results, ablates the leading methods to find effective mechanisms, combines those mechanisms, assesses robustness to \gls{snr} and memory size, and compares computational costs. Section~VI concludes by discussing the implications for the design of \gls{cl} systems in this domain.

%% file: 02_Related_Work/Related_Work.tex
\section{RELATED WORK}
\input{02_Related_Work/Radar_Waveform_Classification}

%% file: 02_Related_Work/Radar_Waveform_Classification.tex
This work focuses on the intersection between radar waveform recognition and \gls{cl}. As only a small number of papers explore this intersection directly, the literature surveyed here encompasses the broader area of radar waveform classification, including works that do not address \gls{cl} explicitly. Recent work in this space spans modulation classification~\cite{2017_Wang_Auto-RWR, rc_jagannath2021, rc_Huang2023, rc_radyolo2024}, transfer learning~\cite{jing2022radar_fewshot, xiao2020radar_fewshot, huang2025radar_fewshot}, and \gls{cl}~\cite{xu2016_radar_il, 2023_il-prototype_rei, 2024_il-amc, 2026_Rabie_ELC}.

Early work on automated waveform and modulation classification demonstrates that \glspl{cnn} can achieve strong performance when applied to \gls{iq}-derived representations. Authors in \cite{2017_Wang_Auto-RWR} convert \gls{iq} data into time-frequency images for \gls{cnn} classification, whilst \cite{rc_jagannath2021} propose a framework for simultaneous modulation classification and signal characterisation, later extended in \cite{rc_Huang2023} via an attention mechanism. A parallel line of work addresses the efficiency of learning new waveforms through transfer learning: \cite{xiao2020radar_fewshot} mitigate overfitting on small-scale radar datasets by reusing prior knowledge, \cite{jing2022radar_fewshot} propose an adaptive loss to transfer knowledge towards difficult waveforms, and \cite{huang2025radar_fewshot} extend these ideas to a self-supervised few-shot regime capable of learning from partially labelled data. Whilst these works collectively establish that prior knowledge can accelerate waveform acquisition, neither line addresses the \gls{cl} setting: the classification works assume a fixed, closed set of waveform classes, and transfer learning alone sacrifices previously learned representations, leaving catastrophic forgetting unresolved.

Several works have applied \gls{cl} directly to radar waveform recognition, though evaluations remain limited in both scale and breadth of benchmarked methods. In \cite{xu2016_radar_il} an incremental learning algorithm is proposed for radar emitter identification in domain-, class-, and task-incremental scenarios, but relies on expert-extracted features that do not scale to large waveform vocabularies. This limitation is addressed in \cite{2023_il-prototype_rei} with a \gls{cnn}-based exemplar framework that freezes feature extraction after initial training and represents new emitters as prototypes in a discriminative feature space shaped by ArcFace loss \cite{2018_ArcFace-Loss}, appending them to an unbiased cosine similarity classifier without retraining. In \cite{2024_il-amc}, a multi-modal framework is proposed for modulation recognition that combines a knowledge distillation penalty with an output-correction layer to mitigate the prediction bias introduced by class imbalance between old exemplars and new data, benchmarking against iCaRL~\cite{ll_mof_2017} and BiC~\cite{2019_bic}. Most recently, \cite{2026_Rabie_ELC} employ a mask-based \gls{cl} method that eliminates inter-task interference whilst permitting forward transfer, evaluated on two synthetic radar datasets.

Across these works, no study considers more than 13 waveform classes or benchmarks more than five \gls{cl} algorithms, so the relative merits of forgetting-mitigation strategies in the radar domain remain unclear. No prior work evaluates both the \gls{til} and \gls{cil} settings, or restricts training to a single pass over each task, as when intercepts arrive as a stream that cannot be stored and revisited. This work evaluates 18 \gls{cl} algorithms (15 under \gls{cil}), spanning replay~\cite{ll_exp_replay_2019, ll_gem_2017, ll_mof_2017}, regularisation~\cite{ll_lwf_2017,ll_ewc_2017,2017_si,2018_rwalk}, optimisation~\cite{ll_gem_2017, AGEM, ll_lamaml_2020}, architectural~\cite{ll_packnet_2018, 2018_hat} and hybrid~\cite{ll_bcl_2020, ctn_2021} approaches. They are benchmarked on \num{54} waveforms from four public datasets~\cite{ds_deeprad, rc_Huang2023, ds_radnist, 2025_White_RadarML}, split into ten tasks, in both settings and under both the \gls{1e} and \gls{5e} regimes.

%% file: 03_Preliminaries/Preliminaries.tex
\section{CONTINUAL LEARNING METHODS}
\input{03_Preliminaries/Continual_Learning_Approaches}

%% file: 03_Preliminaries/Continual_Learning_Approaches.tex
\begin{table*}[t]
    \centering
    \small
    \caption{%
      Summary of \gls{cl} families by mechanism, key strength, and primary limitation.
    }
    \label{tab:cl-approaches}
    \setlength{\tabcolsep}{5pt}
    \renewcommand{\arraystretch}{1.15}
    \begin{spacing}{0.8}
    \resizebox{2\columnwidth}{!}{%
    \begin{tabular}{p{3.5cm} p{5cm} p{4.0cm} p{4.2cm} c}
    \hline\toprule
    \textbf{Approach} & \textbf{Mechanism} & \textbf{Key Strength} & \textbf{Primary Limitation} \\
    \hline\toprule
    \multirow{3}{3.5cm}{\textbf{Regularisation} (LwF\cite{ll_lwf_2017}, EWC\cite{ll_ewc_2017}, SI\cite{2017_si}, RWalk\cite{2018_rwalk}, UCL\cite{ll_ucl_2019})}
      & Penalises changes to parameters identified as important for prior tasks in the parameter or feature space
      & No exemplar storage required; potentially lightweight and broadly applicable across architectures
      & Importance estimation must be reliable and extensible to new tasks \\
    \midrule
    \multirow{2}{3.5cm}{\textbf{Replay} (Res-ER\cite{ll_exp_replay_2019}, Ring-ER\cite{ll_mof_2017}, iCaRL\cite{ll_mof_2017})}
      & Interleaves stored exemplars with current training data via cross-entropy replay
      & Simple and effective; effectiveness scales naturally with buffer size
      & Buffer storage requirement grows with the number of tasks and classes \\
    \midrule
    \multirow{3}{3.5cm}{\textbf{Optimisation} \\ (GEM \cite{ll_gem_2017}, AGEM\cite{AGEM}, C-/S-/La-MAML \cite{ll_lamaml_2020})}
      & Uses stored exemplars or a replay buffer to constrain or shape the gradient update direction via an explicit optimisation procedure
      & Principled gradient-level forgetting prevention without relying solely on rehearsal frequency
      & Constraint projection or bilevel updates introduce significant per-step compute overhead \\
    \midrule
    \multirow{3}{3.5cm}{\textbf{Architectural} \\ (PackNet \cite{ll_packnet_2018}, HAT \cite{2018_hat})}
      & Allocates disjoint parameter subsets per task via pruning or learned masks, preventing any cross-task parameter interference
      & Eliminates forgetting by ensuring that previously learned tasks are unaffected by subsequent tasks
      & Fixed network capacity imposes a hard limit on the number of disjoint subnetworks. \\
    \midrule
    \multirow{3}{3.5cm}{\textbf{Hybrid} \\(CTN \cite{ctn_2021}, BCL-Dual \cite{ll_bcl_2020}, La-ER \cite{ll_lamaml_2020})}
      & Combines replay and optimisation-based updates that decouple stability and plasticity signals across separate memory streams
      & Complementary mechanisms can yield better stability--plasticity trade-offs than either alone
      & Increased architectural and training complexity relative to single-mechanism approaches \\
    \bottomrule\hline
    \end{tabular}%
    }
    \end{spacing}
    \end{table*}
    
    This section explores five families of \gls{cl} methods, grouped by their forgetting-mitigation mechanism: replay, regularisation, optimisation, architectural, and hybrid. Table~\ref{tab:cl-approaches} summarises their relative strengths and limitations. Drawn from these families, the \num{18} methods (\num{15} under CIL) evaluated in later sections are detailed in terms of their individual constituent mechanisms. 
    
    \subsection{Replay-based Methods}
    \label{sect_ll_replay}
    
    \gls{er}-based methods mitigate \gls{cf} by interleaving training with samples drawn from a fixed-size memory of prior observations, approximating the joint training signal.
    
    \gls{er}~\cite{ll_exp_replay_2019} maintains a replay buffer $\mathcal{M}$ of capacity $K$ and minimises cross-entropy over a joint batch combining current-task data with stored exemplars. Two variants are evaluated: Res-ER uses reservoir sampling~\cite{ll_reservoir_sampling} to maintain a global buffer that is asymptotically uniform over all data seen so far, while Ring-ER~\cite{ll_mof_2017} partitions the buffer evenly across tasks, guaranteeing equal per-task retention. Ring-ER therefore benefits from balanced allocation under \gls{til}, while the global objective of Res-ER offers stronger cross-task discrimination under \gls{cil}.
    
    iCaRL~\cite{ll_mof_2017} combines exemplar replay with knowledge distillation. Exemplars are selected by herding to approximate each class mean in feature space, with the fixed budget $K$ redistributed uniformly as new classes arrive. The cross-entropy loss is augmented with a \gls{kl} distillation penalty against the model frozen at each task boundary,
    \begin{equation}\label{eqn:icarl_loss}
        \mathcal{L}_\text{KL} = \lambda \sum_{\tau=0}^{t-1}
          D_\text{KL}\!\left(
            \sigma\!\left(f_{\theta^*_\tau}(x^\tau)\right)
            \,\Big\|\,
            \sigma\!\left(f_\theta(x^\tau)\right)
          \right),
    \end{equation}
    where $\sigma(\cdot)$ is the sigmoid, $f_{\theta^*_\tau}$ is the model frozen at the end of task $\tau$, and $\lambda$ controls the distillation strength. At inference, a test sample is assigned to the nearest class prototype under cosine similarity.
    \subsection{Regularisation-based Methods}
    \label{sect_ll_reg}
    
    Regularisation-based methods mitigate \gls{cf} by penalising destructive parameter updates, either in weight space via importance-weighted parameter drift, or in function space by constraining the model's outputs to remain consistent with those of prior tasks.
    
    \noindent\textbf{Function Regularisation.} LwF~\cite{ll_lwf_2017} applies the function-space variant: a frozen snapshot $f_{\theta^*}$ of the model after each task boundary serves as a teacher, and the student is trained on current-task labels plus a similar $\mathcal{L}_\text{KL}$ term to iCaRL~\eqref{eqn:icarl_loss}, applied to the current data stream using a single teacher. LwF requires no episodic memory.
    
    \noindent\textbf{Parameter Regularisation.} EWC~\cite{ll_ewc_2017}, SI~\cite{2017_si}, and RWalk~\cite{2018_rwalk} share a common consolidation objective \eqref{eqn:param_importance_generic}, differing only in how the importance $I_i$ is estimated. EWC sets $I_i$ to the diagonal of the Fisher information matrix, approximating the \gls{kl} cost of drifting from the old-task optimum. SI instead accumulates importance online as a path integral of each parameter's contribution to loss reduction along the optimisation trajectory. RWalk combines an \gls{ema} of the diagonal Fisher with an SI-style path statistic normalised by the Fisher metric, yielding $I_i = F_i + S_i$.
    \begin{equation}\label{eqn:param_importance_generic}
        \mathcal{L} = \mathcal{L}_\text{CE} + \lambda \sum_i I_i\,(\theta_i - \theta_i^*)^2,
    \end{equation}
    
    UCL~\cite{ll_ucl_2019} extends parameter importance to a fully Bayesian form, replacing deterministic weights with factorised Gaussian posteriors and penalising drift in both the mean and the variance of each weight. The mean penalty is an importance-weighted analogue of~\eqref{eqn:param_importance_generic}, with importance derived from the ratio of initial to current posterior uncertainty, while a log-corrected variance penalty makes sharpening the posterior cheaper than broadening it. Collectively these terms approximate the \gls{kl} divergence between successive weight posteriors.
    \subsection{Optimisation-based Methods}
    \label{subsec:lamaml}
    
    \noindent\textbf{Gradient Projection.} GEM~\cite{ll_gem_2017} constrains each update so that the loss on every previously observed task cannot increase, accepting the gradient only if it has non-negative inner product with every prior-task reference gradient and otherwise projecting it via a \gls{qp}; the \gls{qp} grows with the number of tasks, making GEM expensive on long sequences. AGEM~\cite{AGEM} replaces the per-task constraint with an averaged constraint, projecting the gradient $g$ if $\langle g,g_\text{ref}\rangle {<}0$ for reference gradient $g_\text{ref}$ drawn from all prior buffers \eqref{eqn:agem_projection}, reducing per-step overhead to a single inner product while trading away GEM's per-task guarantee.
    \begin{equation}\label{eqn:agem_projection}
        \tilde{g} = g - \frac{\langle g, g_\text{ref} \rangle}{\langle g_\text{ref}, g_\text{ref} \rangle} \, g_\text{ref},
    \end{equation}
    
    \noindent\textbf{Meta-Learning.} Meta-learning methods frame continual adaptation as a bilevel optimisation problem, meta-training an initialisation from which rapid, low-forgetting adaptation to new tasks is possible. La-MAML~\cite{ll_lamaml_2020} adapts the MAML~\cite{2017_maml} framework to \gls{cl} by replacing its held-out outer-loop data with a reservoir-sampled replay buffer: an inner loop produces fast weights $\theta'$ via a gradient step scaled by per-parameter rates $\hat{\alpha}_j$,
    \begin{equation}\label{eqn:lamaml_inner}
        \theta'_j = \theta_j - \hat{\alpha}_j \cdot \frac{\partial \mathcal{L}_\text{CE}(B_i;\,\theta)}{\partial \theta_j},
    \end{equation}
    and the outer loop evaluates these fast weights against replay-augmented meta-batches to form a meta-loss,
    \begin{equation}\label{eqn:lamaml_meta_loss}
        \mathcal{L}_\text{meta}
        =
        \frac{1}{K_\text{meta}}
        \sum_{i=1}^{K_\text{meta}}
        \mathcal{L}_\text{CE}\!\left(B^\text{rep};\, \theta'_i\right),
    \end{equation}
    which first updates the rates and then updates the backbone using those rates as step sizes, $\theta_j \leftarrow \theta_j - \hat{\alpha}_j\,\partial \mathcal{L}_\text{meta}/\partial \theta_j$; here $B^\text{rep}$ is a replay batch, $\theta'_i$ the fast weights adapted on the $i$-th inner batch, and $K_\text{meta}$ the number of meta-batches averaged. CMAML and SMAML ablate this scheme: CMAML fixes the rates at $\alpha_\text{init}$ and updates the backbone with a single global rate $\beta$ in place of $\hat{\alpha}_j$, while SMAML meta-trains the rates as in La-MAML but still applies CMAML's scalar update, so the updated rates take effect only at the next inner step. This contrast isolates the effect of per-parameter, rather than scalar, learning-rate adaptation.
    
    \subsection{Architectural Methods}
    \label{subsec:packnet}
    
    Architectural methods prevent \gls{cf} by allocating disjoint parameter subsets per task, so that learning a new task cannot alter representations learned for previous ones. PackNet~\cite{ll_packnet_2018} achieves this via iterative magnitude-based pruning. After each task, the largest-magnitude fraction $\rho$ of currently free weights is permanently locked to that task and restored after every optimiser step, confining subsequent plasticity to the remaining pool. The free pool shrinks geometrically; earlier tasks receive more parameters than later ones. At inference, a task-aware binary mask routes the forward pass through the corresponding sub-network. 
    
    HAT~\cite{2018_hat} instead learns soft attention masks end-to-end: a task-specific embedding is passed through a sigmoid scaled by an annealing parameter $s$,
    \begin{equation}\label{eqn:hat_gate}
        \mathbf{m}_l^{(t)} = \sigma\bigl(s\,\mathbf{e}_l^{(t)}\bigr) \in (0,1)^{C},
    \end{equation}
    producing a gate over the layer's channels that hardens towards binary values as $s$ is annealed. At each task boundary the gates are binarised into a mask that zeroes gradients on previously allocated weights for all subsequent tasks, with an added regularisation term discouraging the current task from claiming unused capacity unnecessarily.
    
    \subsection{Hybrid Methods}
    \label{sec:hybrid}
    
    Hybrid methods combine mechanisms from the preceding categories within a bilevel optimisation scheme. The scheme has two levels, each a separate parameter set or update loop, and each learns from its own data stream. This division supports task-specific adaptation at the inner level while the outer level preserves cross-task generalisation. Both hybrids evaluated here train their inner level on a common replay-augmented objective,
    \begin{equation}\label{eqn:hybrid_inner}
    \begin{split}
        \mathcal{L}^{\text{train}}
        =\;&
        \mathcal{L}_{\text{CE}}(x, y)
        + \lambda_{\text{rep}}\, \mathcal{L}_{\text{CE}}(x^{\text{rep}}, y^{\text{rep}}) \\
        &+ \lambda_{\text{KL}}\,
          D_{\text{KL}}\!\left( 
              \sigma\!\left(\tfrac{f_{\theta^*_\tau}(x^{\text{rep}})}{\tau}\right)
              \;\Big\|\; 
              \sigma\!\left(\tfrac{f(x^{\text{rep}})}{\tau}\right)
          \right),
    \end{split}
    \end{equation}
    an instance of the distillation penalty~\eqref{eqn:icarl_loss} whose soft labels are recorded at the previous task boundary $\theta^*_\tau$.
    
    CTN~\cite{ctn_2021} splits the model into a shared backbone $\phi$ and a controller $\theta$. The controller maps task embeddings to per-task \gls{film} parameters $\{\gamma_t,\beta_t\}$ that modulate the backbone features,
    \begin{equation}\label{eqn:ctn_film}
        \tilde{h}(x;t) = \frac{\gamma_t}{\|\gamma_t\|_2} \odot \hat{h}(x) + \frac{\beta_t}{\|\beta_t\|_2}.
    \end{equation}
    The backbone trains on the replay-augmented objective~\eqref{eqn:hybrid_inner}. The controller trains on a separate stream, which encourages transformations that generalise beyond the immediate distribution. BCL~\cite{ll_bcl_2020} instead operates over a single shared parameter vector. A working memory supplies the same replay-augmented objective~\eqref{eqn:hybrid_inner} to an inner loop, while a separate meta memory provides clean supervision to a single outer step. BCL folds the resulting adapted state $\phi'$ back via Reptile-style interpolation~\cite{2018_reptile},
    \begin{equation}\label{eqn:bcl_interp}
        \theta \leftarrow \theta + \beta\left(\phi' - \theta\right).
    \end{equation}
    La-ER~\cite{ll_lamaml_2020}, an ablation of La-MAML, isolates the meta-learned step sizes from the MAML-style update. The per-parameter rates are still meta-trained, but the backbone is updated by plain replay rather than by the meta-loss.

%% file: 04_Methodology/Methodology.tex
\section{METHODOLOGY} \label{sec:methodology}
\input{04_Methodology/Problem_Formulation}
\input{04_Methodology/Performance_Evaluation}
\input{04_Methodology/Datasets}

\input{04_Methodology/Experimental_Setup}
\input{04_Methodology/Ablation_Protocol}

%% file: 04_Methodology/Problem_Formulation.tex
\subsection{Problem Formulation}
\subsubsection{Base Learning Objective}
The base learning objective for all methods is the cross-entropy loss between the predicted class distribution and the ground-truth label. Let $\mathbf{x} {\in} \mathcal{X}$ denote an input \gls{iq} sample and $y {\in} \mathcal{Y}_t$ its class label. The predicted class probability is $p(y {\mid} \mathbf{x}){=}[\text{softmax}(f_{\boldsymbol{\theta}}(\mathbf{x}))]_y$, and the class-weighted cross-entropy over a batch of $N$ samples is given by \eqref{eq:weighted_ce}, where $w_{y_i}$ is a per-class weight that addresses class imbalance. The loss is normalised by the total weight in the batch.
\begin{equation} \label{eq:weighted_ce}
    \mathcal{L}_\text{CE}{=}-\frac{\sum_{i=1}^{N} w_{y_i} \log p(y_i \mid \mathbf{x}_i)}
                                 {\sum_{i=1}^{N} w_{y_i}}
\end{equation}
Let $\mathcal{B} {\subseteq} \mathcal{Y}_t$ denote the set of classes present in the batch and $n_c$ the number of samples of class $c$. The weight of class $c$ is its inverse frequency, normalised so that the mean weight over observed classes is unity, i.e.\ $\frac{1}{|\mathcal{B}|}\sum_{c \in \mathcal{B}} w_c{=}1$:
\begin{equation} \label{eq:class_weights}
    w_c{=}\dfrac{N / n_c}{\dfrac{1}{|\mathcal{B}|}\displaystyle\sum_{c' \in \mathcal{B}} N / n_{c'}},
    \quad c \in \mathcal{B}
\end{equation}
\subsubsection{Continual Learning Objective} \label{sec:cl_objective}
A model learns a sequence of $T$ tasks, $\mathcal{D}{=}\{\mathcal{D}_1, \dots, \mathcal{D}_T\}$. Task $t$ is defined by a dataset $\mathcal{D}_t{=}\{(\mathbf{x}_i^t, y_i^t)\}_{i=1}^{N_t}$ of $N_t$ \gls{iq} samples with labels $y_i^t \in \mathcal{Y}_t$, and class sets are disjoint across tasks ($\mathcal{Y}_i {\cap} \mathcal{Y}_j{=}\emptyset$ for $i {\neq} j$). During task $t$, only $\mathcal{D}_t$ is observed; prior data are available only if retained by a replay method.

The model $f_{\boldsymbol{\theta}}{=}h_{\boldsymbol{\psi}} \circ g_{\boldsymbol{\phi}}$ comprises a shared feature extractor $g_{\boldsymbol{\phi}}\!:\! \mathcal{X} \!\rightarrow\! \mathcal{Z}$ and a classifier head $h_{\boldsymbol{\psi}}\!:\! \mathcal{Z} \!\rightarrow\! \mathbb{R}^{C}$, with $\boldsymbol{\theta}{=}\{\boldsymbol{\phi}, \boldsymbol{\psi}\}$. Under \gls{til}, a \gls{tid} is given at inference and selects a task-specific head $\boldsymbol{\psi}_t$, so $C{=}|\mathcal{Y}_t|$. Under \gls{cil}, no \gls{tid} is given, and a single shared head covers every class seen so far, so $C{=}|\mathcal{Y}_{1:t}|$ with $\mathcal{Y}_{1:t}{=}\bigcup_{\tau=1}^{t} \mathcal{Y}_\tau$.

At task $t$, the ideal objective is the cumulative risk over all tasks observed so far \eqref{eq:cl_objective}, where $\mathcal{L}_\tau$ denotes the loss~\eqref{eq:weighted_ce} on $\mathcal{D}_\tau$. Only the term $\mathcal{L}_t$ can be evaluated, since $\mathcal{D}_\tau$ is unavailable for $\tau < t$. The methods considered here differ in how they approximate the remaining terms.
\begin{equation} \label{eq:cl_objective}
    \min_{\boldsymbol{\theta}} \sum_{\tau=1}^{t} \mathcal{L}_\tau(\boldsymbol{\theta}),
\end{equation}

%% file: 04_Methodology/Performance_Evaluation.tex
\subsection{Performance Evaluation}
\subsubsection{Classification of Radar Waveforms}
Classification performance of the radar waveforms is evaluated using precision, recall, and F1 score, computed per class and aggregated via macro-averaging across all waveform classes. Macro-averaging treats each class equally regardless of sample count. Per-class precision, recall and F1 scores are defined in \eqref{eq:precision}, where $\text{TP}$, $\text{FP}$, and $\text{FN}$ denote the number of true positives, false positives, and false negatives, respectively.
\begin{equation} \label{eq:precision}
    P = \frac{\text{TP}}{\text{TP} + \text{FP}},\quad R = \frac{\text{TP}}{\text{TP} + \text{FN}},\quad \mathrm{F1} {=} \frac{2PR}{P {+} R}
\end{equation}
Macro-averaged precision, recall and F1 are then defined as \eqref{eq:macro}, where $C$ is the number of classes. 
\begin{equation} \label{eq:macro}
    \bar{P} = \frac{1}{C}\sum_{c=1}^{C} P, \;\; 
    \bar{R} = \frac{1}{C}\sum_{c=1}^{C} R, \;\;
    \overline{\mathrm{F1}} = \frac{1}{C}\sum_{c=1}^{C} \mathrm{F1}
\end{equation}
\subsubsection{Continual Learning}
The degree to which prior task performance is preserved after learning task $t$ is measured by \gls{bwt}. \gls{fwt}~\eqref{eq:fwt} measures how well the model performs on a task before training on it, relative to a random model. BWT and FWT are defined in \eqref{eq:bwt}--\eqref{eq:fwt}, where $\mathrm{F1}_{i,j}$ denotes the F1 on task $j$ after training on task $i$, and $\overline{\mathrm{F1}}_t$ is the F1 of a randomly initialised model on task $t$. A negative \gls{bwt} indicates catastrophic forgetting, and a positive \gls{fwt} indicates prior learning has benefited subsequent tasks. 
\begin{equation} \label{eq:bwt}
    \text{BWT} = \frac{1}{T-1} \sum_{t=1}^{T-1} \left( \mathrm{F1}_{T,t} - \mathrm{F1}_{t,t} \right)
\end{equation}
\begin{equation} \label{eq:fwt}
    \text{FWT} = \frac{1}{T-1} \sum_{t=2}^{T} \left( \mathrm{F1}_{t-1,t} - \overline{\mathrm{F1}}_t \right)
\end{equation}
Overall model performance, $\overline{\mathrm{F1}}_\text{CL}$, is defined as the class-weighted mean of $\overline{\mathrm{F1}}$ across all $T$ tasks \eqref{eq:f1_cl}, where $n_c^t$ is the number of classes in task $t$ and $\overline{\mathrm{F1}}^{(T,t)}$ denotes $\overline{\mathrm{F1}}$ on task $t$ after training on all tasks. Under CIL, $\fcl$ is the F1 score over all seen classes.
\begin{equation} \label{eq:f1_cl}
    \mathrm{F1}_\text{CL} = \frac{1}{\sum_t^T n_c^t} \sum_{t=1}^{T} n_c^t\,\overline{\mathrm{F1}}^{(T,t)}
\end{equation}
CL methods are evaluated against two reference models. The \gls{ft} model trains on each task without any mitigation against forgetting, giving a practical lower bound. The \gls{jt} model trains on the data of all tasks so far, jointly optimising them with the current task, and gives a practical upper bound.

%% file: 04_Methodology/Datasets.tex
\subsection{Datasets}
\subsubsection{Description}
Four datasets are used in this work, all of which are synthetic and one contains \gls{ota} data (synthetic data transmitted and then received on hardware). Table~\ref{tab:signal_types_in_datasets} summarises the signal types within each dataset.

\textbf{\gls{rml}} \cite{2025_White_RadarML}
is an \gls{ota} synthetic dataset containing 15 signal types. Signals are synthetically generated with one of 15 modulation scheme and phase-coding combinations: Barker (5-13, 5$\times$11, 5$\times$13), \gls{cw}, \gls{fmcw}, \gls{fsk} (0-3), \gls{hfm}, \gls{qfm} and \gls{nlfm}. These are then transmitted and received by a Radio-Frequency-System-on-a-chip (RFSoC) on different channels. Three channels receive the data simultaneously in an attempt to isolate channel distortions from the transmission. The sampling rate is $3.84~\text{GS/s}$, downsampled to $120~\text{MS/s}$. The synthetic waveforms are generated and transmitted at high \gls{snr}. 

\textbf{\gls{rnist}} \cite{ds_radnist} contains 
five types of pulses corresponding to radar working modes in the 3.55–3.7~GHz band \cite{ds_radnist}. It contains two unmodulated (P0N) and three frequency/angle modulated (Q3N) pulse types. All pulses vary in pulse width ($0.5{-}100~\mu\text{s}$), \gls{pri} ($0.3{-}3~\text{ms}$) and \gls{ppb} ($2{-}40$). The parameters for each pulse are randomly sampled per burst. Additionally, P0N pulses vary in phase-coding (Barker, Frank, P1-4/x, Zadoff-Chu), and Q3N in bandwidth ($1{-}100~\text{MHz}$). The sampling rate is $10~\text{MS/s}$.

\textbf{\gls{rc}} \cite{rc_Huang2023} contains radar pulses according to five modulation schemes: coherent pulse train, Barker code, polyphase Barker code, Frank code and linear frequency modulation. Pulse parameters vary in code length, pulse width, \gls{pri} and \gls{ppb}. The sampling rate is $3.2$~MS/s.

\textbf{\gls{dr}} \cite{ds_deeprad} contains 22 classes of pulses generated according to modulation and phase-code combinations, including non-modulation. The modulation schemes used are: \gls{lfm}, \gls{fsk} (1-3 bits/symbol), \gls{psk} (1-3 bits/symbol), \gls{pm} (Barker, Frank, P1-4/x, Zadoff-Chu, T1-4), and non-modulation. The sampling rate is $100~\text{MS/s}$.

\begin{table}[t]
\centering
\caption{Signal types available in evaluated datasets.}
\label{tab:signal_types_in_datasets}
\resizebox{0.95\columnwidth}{!}{%
\begin{tabular}{>{\raggedright\arraybackslash}m{4cm}cccc}
\hline\toprule
\textbf{Signal Type / Family}           & \textbf{DeepRad} & \textbf{RadNIST} & \textbf{RadChar} & \textbf{RadarML} \\
\hline\toprule

Non-modulation                  & \checkmark & \checkmark &            & \checkmark \\ \midrule

FM (LFM, FMCW, NLFM, HFM, QFM)           & \checkmark & \checkmark & \checkmark & \checkmark \\ \midrule

Coherent pulse train                     &            &            & \checkmark &  \\ \midrule

FSK (2/4/8-FSK)                          & \checkmark &            &            & \checkmark \\ \midrule

PSK (2/4/8-PSK)                          & \checkmark &            &            &  \\ \midrule

Costas code                              & \checkmark &            &            &  \\ \midrule

Barker codes                             & \checkmark & \checkmark & \checkmark & \checkmark \\ \midrule

Polyphase Barker                         &            &            & \checkmark &  \\ \midrule

Frank code                               & \checkmark & \checkmark & \checkmark &  \\ \midrule

Huffman code                             & \checkmark &            &            &  \\ \midrule

P1–P4 \& Px codes              & \checkmark & \checkmark &            &  \\ \midrule

Zadoff–Chu                               & \checkmark & \checkmark &            &  \\ \midrule

T1–T4 phase codes                        & \checkmark &            &            &  \\ \midrule 

Q3N radar modes                          &            & \checkmark &            &  \\ \midrule

P0N radar modes                          &            & \checkmark &            &  \\ 
\bottomrule\hline
\end{tabular}%
}
\end{table}

\subsubsection{Data Processing}
For all but the \gls{rml} dataset, the only processing steps required are i) filtering the SNR range to $-10{-}10\text{dB}$ for consistency, and ii) matching the dimensions of the waveform slices, across datasets, to be fed as inputs to the backbone ResNet. The common SNR range chosen is $-10{-}10\text{dB}$; and the sample dimensions chosen are $[2, 512]$, corresponding to two (I/Q) channels each with $512$ samples. \gls{rml} requires additional processing steps: waveform detection and extraction, inter-ADC phase alignment, and linear combination of three ADC channels.

Each RadarML recording is $140\,\mu\text{s}$ in duration and sampled at 
$120\,\text{MS/s}$. However, the waveforms have a much smaller pulse width of $10{-}100~\mu\text{s}$. Waveforms are thus extracted using a threshold-based edge-detection algorithm. The algorithm locates the waveform edges and crops a window around it. Each window has length $n L$, where $n\in\mathbb{Z}^+$ and $L$ is the model's input length. Each window is divided into $n$ disjoint input slices, each receiving the same class label. 

Each waveform in RadarML is captured simultaneously across three ADC channels.
Prior to use, inter-ADC phase misalignment is corrected by aligning channels~2
and~3 to channel~1, which serves as the reference $\mathbf{x}_\text{ref}$.
The phase offset of a channel $\mathbf{x}$ relative to the reference is
estimated from their complex inner product,
$\rho = \mathbf{x}_\text{ref}^{H}\mathbf{x}$, and alignment is applied as $\mathbf{x}_\text{aligned} = \mathbf{x}\, e^{-j \arg(\rho)}$.
\noindent Following alignment, a similarity score is computed for each pair of
channels \eqref{eq:cos_similarity}. Samples for which any pairwise score falls
below $0.9$ are discarded as insufficiently consistent, accounting for
approximately $4.5\%$ of the dataset. \gls{awgn} is then superimposed onto the
remaining samples to bring them within the target range of $-10$ to $10$~dB.
\begin{equation} \label{eq:cos_similarity}
    \text{sim}(\mathbf{A}, \mathbf{B}) =
    \frac{\operatorname{Re}\!\left(\mathbf{A}^{H}\mathbf{B}\right)}
         {\|\mathbf{A}\|\,\|\mathbf{B}\|}
\end{equation}
The corrected 3-ADC \gls{iq} waveforms are subsequently reduced to the two-channel format expected by the backbone via a learnable linear layer. This layer learns a linear transformation of the three ADC channels, initialised to a uniform average.

\subsection{Task Construction}
\label{sec:task-construction}

Table~\ref{tab:signal_types_in_datasets} shows several modulations, such as Barker codes, that appear in more than one dataset. These are shared modulations, not shared \emph{waveform classes}. Overlap would matter if the model classified modulation alone. In this work, however, each class is the combination of an underlying waveform \emph{and} its modulation. The analysis in Appendix~\ref{apx:overlapping-mods} tests this by measuring how often a model confuses same-modulation waveforms, relative to chance. No duplicate waveforms were found.

Tasks are constructed from the four datasets. For each dataset, the set of signal classes is randomly partitioned into disjoint groups, with each group defining one task. The number of slices per class is fixed to $10\text{k}$ for all datasets. The resulting assignments are summarised in Table~\ref{tab:class-splits}, alongside task difficulty measured as F1.

Since each task is drawn from a single dataset, differences between datasets (e.g. sampling rate) could leak \gls{tid}. Appendix~\ref{apx:tid-leakage} finds that models retain a significant amount of \gls{tid}. However, this retention correlates strongly with task difficulty (Spearman $\rho{=}0.84$), which suggests that it is not primarily caused by \gls{tid} leakage.
\begin{table}[t]
\caption{Class partitions per task across datasets. Size indicates number of samples before training/testing split. Difficulty is the F1 (\%) of a model trained on that task alone for $5$ epochs; lower values indicate harder tasks.}
\label{tab:class-splits}
\centering
\resizebox{1\columnwidth}{!}{%
\begin{tabular}{lc>{\raggedright\arraybackslash}p{8cm}ccc}
\hline\toprule
\textbf{Dataset} & \textbf{Task} & \textbf{Signal Types (Classes)} & \textbf{\# Classes} & \textbf{Size} & \textbf{Difficulty} \\
\hline\toprule
\gls{rc}  & 1 & Coherent Pulse Train, Barker Code, Polyphase Barker Code, Frank Code, \gls{lfm} & 5 & 50k & \pmv{97.4}{0.1} \\
\midrule
\gls{rnist} & 1 & Q3N\#2, Q3N\#1, Q3N\#3, P0N\#2-P3, P0N\#2-Frank, P0N\#2-Px & 6 & 60k & \pmv{90.1}{0.4} \\
\gls{rnist} & 2 & P0N\#1, P0N\#2-Barker, P0N\#2-P2, P0N\#2-P4, P0N\#2-P1, P0N\#2-Zadoff-Chu & 6 & 60k & \pmv{96.0}{0.2} \\
\midrule
\gls{rml} & 1 & \gls{fsk}\#0, Barker-5, Barker-13, Barker-11, \gls{fmcw} & 5 & 50k & \pmv{81.7}{0.9} \\
\gls{rml} & 2 & Barker-5${\times}$11, Barker-5${\times}$13, Barker-7, \gls{fsk}\#3, \gls{cw} & 5 & 50k & \pmv{84.5}{0.4} \\
\gls{rml} & 3 & \gls{fsk}\#1, \gls{qfm}, \gls{hfm}, \gls{fsk}\#2, \gls{nlfm} & 5 & 50k & \pmv{73.9}{1.2} \\
\midrule
\gls{dr} & 1 & 8-\gls{fsk}, Barker, T4, T1, Huffman, Px & 6 & 60k & \pmv{66.6}{1.1} \\
\gls{dr} & 2 & 2-\gls{psk}, P3, \gls{nlfm}, \gls{lfm}, P2, Zadoff-Chu & 6 & 60k & \pmv{75.0}{2.1} \\
\gls{dr} & 3 & P1, 4-\gls{fsk}, 8-\gls{psk}, Frank, 4-\gls{psk} & 5 & 50k & \pmv{57.4}{1.5} \\
\gls{dr} & 4 & FM-Costas, T2, P4, 2-FSK, T3 & 5 & 50k & \pmv{59.8}{1.7} \\
\bottomrule\hline
\end{tabular}%
}
\end{table}

%% file: 04_Methodology/Experimental_Setup.tex
\subsection{Backbone and Hyperparameters}
The backbone is a 1D ResNet18~\cite{2016_He_ResNet}, obtained by replacing every 2D operation with its 1D counterpart. Batch normalisation~\cite{2015_BatchNorm} is retained. Table~\ref{tab:shared-hyperparams} lists the hyperparameters shared by all methods.

All experiments use \gls{sgd}~\cite{2010_sgd} with momentum $0.9$ and no weight decay. Adaptive optimisers such as Adam~\cite{2014_adam-optimiser} are avoided because their per-parameter step sizes interact with the \gls{cl} mechanism under test. With \gls{sgd}, differences in performance can be attributed to the \gls{cl} methods themselves.

Following~\cite{AGEM}, hyperparameters are tuned on a held-out split formed by randomly sampling $15\,\%$ of each source dataset. These samples are excluded from all training and evaluation. The tuning split comprises four tasks, one per source dataset, of five classes each. Each task is trained for five epochs. The class-to-task partition is different from the main experiments so that tuning does not favour a particular task partition. The \gls{lr} is tuned first on the \gls{jt} model and then fixed for all methods. The remaining hyperparameters are tuned hierarchically in estimated order of importance. Appendix~\ref{sec:appendixA} lists the sweep ranges and selected values.
\begin{table}
\centering
\caption{Shared hyperparameters. Replay parameters are only applicable to methods that use a memory buffer.}
\label{tab:shared-hyperparams}
\resizebox{0.8\columnwidth}{!}{%
\begin{tabular}{p{5cm}|p{5cm}}
\hline\toprule
\textbf{Hyperparameter} & \textbf{Value} \\
\hline\toprule
Architecture & ResNet18-1D \\
Optimiser & SGD \\
Epochs &  1 (single-epoch) / 5 (multi-epoch) \\
Batch Size & 256 \\
Replay Batch Size & 128 \\
Replay Memory Size &  5120 \\
Batch Scaling & Normalisation \\
Training/Testing Split & 80-20 \\
\bottomrule\hline
\end{tabular}%
}
\end{table}

%% file: 04_Methodology/Ablation_Protocol.tex
\subsection{Ablations and Significance Testing}
\label{sec:stat_protocol}
The leading methods each bundle several mechanisms (episodic replay, gradient projection, meta-learned updates, bilevel optimisation, distillation, task conditioning), so $\fcl$ alone does not reveal which of them carries the performance. Leave-one-out ablations isolate each one: every ablation removes or alters a single mechanism and re-runs the full experiment in both training regimes. Optimisation budget (SGD steps per batch) is matched wherever an ablation would otherwise change it. Mechanisms that prove significant are then combined into new models, each built on one of the leading methods.

Each ablation is compared to its family's full-method baseline under a paired design, so pairing cancels the between-seed variance. Each ablation uses a common seed set, $n{=}6$ by default, and reports the mean effect $\overline{\Delta}$ with a $95\%$ \gls{ci} (Student's $t$) and a Holm--Bonferroni-corrected paired $t$-test ($p_{\text{Holm}}$)~\cite{holm1979}, corrected within each method's ablation family. Following~\cite{lakens2018}, each ablation receives one of four verdicts: \emph{\gls{ss}} ($p_{\text{Holm}}{<}0.05$; \gls{ci} excludes $0$), \emph{\gls{ms}} ($p_{\text{Holm}}{>}0.05$; CI excludes $0$), \emph{\gls{ns}} (CI within $\pm\delta$), and \emph{\gls{inc}} (CI spans $0$ and exceeds $\pm\delta$). Inconclusive ablations are re-run at $n{=}12$ to tighten the \gls{ci}; those that remain inconclusive are reported as such. The \gls{sesoi} is $\delta{=}1\%$ $\fcl$, the median gap between the ablated methods and their neighbours in the main benchmark. Combined models are additionally re-tested on six held-out seeds, disjoint from the ablation grid.

%% file: 05_Results/Results.tex
\section{RESULTS}\label{sec:results}
\subsection{Overview}
Methods are benchmarked on the ten-task radar waveform classification problem in both \gls{til} and \gls{cil} settings, each under a \gls{1e} regime (every task seen once) and a \gls{5e} regime (\num{5} epochs per task). Each setting is reported in three parts: the benchmark itself, the ablations of the leading methods (Section~\ref{sec:stat_protocol}), and robustness to SNR and buffer size. The significant mechanisms are then combined into new models, and the section closes with the memory and runtime cost of every method. Results are averaged over $6$ seeds which also vary task order.

\subsection{Task-Incremental Setting}
\input{05_Results/Tables/til_table}
\begin{figure}
    \centering
    \begin{subfigure}{\linewidth}
        \includegraphics[width=\linewidth]{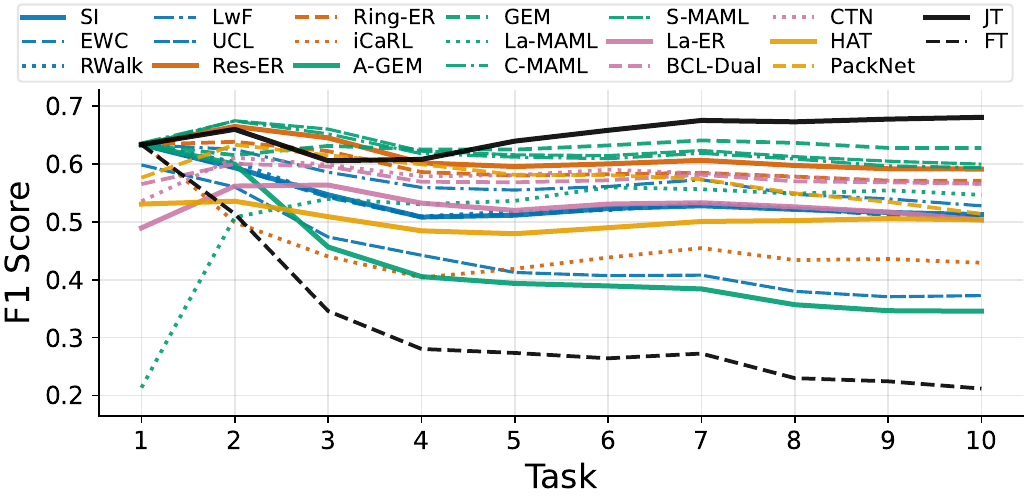}
    \end{subfigure}\\[0em]
    \begin{subfigure}{\linewidth}
        \includegraphics[trim={0cm 0cm 0cm 1.5cm}, clip,width=\linewidth]{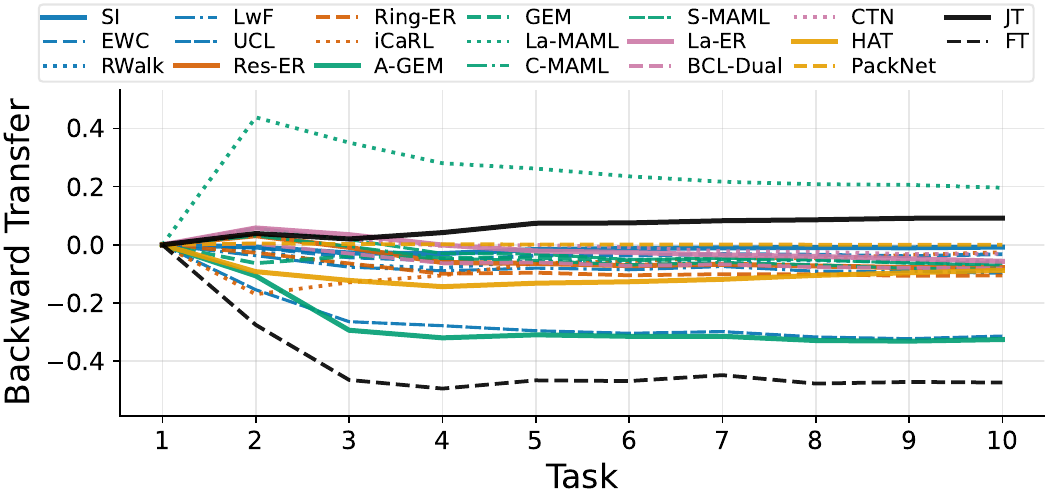}
    \end{subfigure}\\[0em]
    \caption{Per-task \gls{cl} metrics for the \gls{1e} \gls{til} setting: $\bar{F1}_\text{CL}$ (top) and \gls{bwt} (bottom) as tasks are introduced. Both figures share a legend.}
    \label{fig:se_til_per_task_metrics}
\end{figure}
\subsubsection{Benchmark}
Table~\ref{tab:results_til} reports all \num{18} methods. Retained data is the clearest marker of a strong method: eight of the ten leading \gls{1e} methods, and nine of ten in \gls{5e}, store samples from earlier tasks. GEM leads both regimes, closing $88{-}91\%$ of the gap between FT and JT. A-GEM, which collapses GEM's per-task constraints into a single averaged one, is the weakest method in both ($21{-}29\%$), so the approximation is too coarse for these tasks. Among the non-replay methods, EWC, SI and RWalk are indistinguishable from one another, and LwF leads them in both regimes.

The extra epochs help almost every method more than they help either baseline (mean ${+}7.7\%$ against ${+}4.1\%$ for FT and ${+}4.9\%$ for JT); only A-GEM and UCL gain nothing. The gain costs retention for the replay methods (GEM's \gls{bwt} falls from ${-}6.4\%$ to ${-}12.8\%$) but not for the parameter regularisers, whose \gls{bwt} moves towards zero. The field also compresses: the \gls{5e} top six lie within $3.0\%$ of one another.

The learning curves (Fig.~\ref{fig:se_til_per_task_metrics}) show that the ranking is largely settled by Task~3, and separate three failure modes. A-GEM and UCL learn each task but fail to keep it (stability-limited). iCaRL keeps what it learns but learns each task poorly (plasticity-limited), which is why it gains the most from extra epochs (${+}18.2\%$). La-MAML learns Task~1 poorly and acquires it later through replay, so its leading \gls{1e} \gls{bwt} (${+}19.6\%$) reflects late acquisition rather than retention; with five epochs it turns negative.

\input{05_Results/Tables/til_ablations_compact}

\subsubsection{Ablating Methods and Combining SS Mechanisms}
Table~\ref{tab:til_ablations} ablates the five leading methods (GEM, Res-ER, C-MAML, BCL-Dual and CTN) to find out which of their mechanisms are \gls{ss}. The additivity of \gls{ss} mechanisms is then tested by grafting them onto the leading method (GEM) and a lightweight alternative. Four findings emerge.
\begin{enumerate}[leftmargin=*]
    \item \textbf{Retention mechanisms carry the performance.} Replacing GEM's gradient projection with plain replay from the same buffer (G1) is the largest effect in the grid ($-4.8$/$-3.2$). \Gls{kl} distillation is \gls{ss} in both BCL-Dual (B1) and CTN (T2) in both regimes, and its effect roughly doubles with the extra epochs, where replay methods forget more. For Res-ER, a static ring buffer costs $1.5\%$ (E1), but a fully-utilised ring recovers almost all of it (E2). The advantage of Res-ER over Ring-ER is therefore mostly due to buffer utilisation.
    \item \textbf{Optimisation-side mechanisms are suited for limited compute only.} BCL-Dual's bilevel components (B2, B3, B4) and CTN's FiLM conditioning (T1) each cost at most $2.1\%$ when removed in \gls{1e}. However, all but B2 lose significance once more epochs are allowed.  
    \item \textbf{Meta-learning contributes nothing.} Removing C-MAML's inner-loop adaptation (M2) or meta-batch averaging (M1) is \gls{ns} or \gls{ms} in \gls{1e}, and \emph{improves} $\fcl$ in \gls{5e} ($+1.4$ and $+0.8$, both \gls{ss}).
    \item \textbf{Distillation and Dynamic Ring Buffering Improve GEM.} Rows A1--C1 and R1 show that SS mechanisms are additive. Gradient projection combines with KL distillation and dynamic ring buffering. GEM$^+$ reaches $66.2\%$ and $74.2\%$ in \gls{1e} and \gls{5e}, respectively. This closes $96\%$ of the JT--FT gap in \gls{1e}. Res-ER$^+$ is also introduced, as a lightweight alternative to GEM/GEM$^+$ (see Section~\ref{sec:cost}).
\end{enumerate}

\subsubsection{Robustness to SNR and Buffer Size}
\begin{figure}[t]
    \centering
    \includegraphics[width=\linewidth]{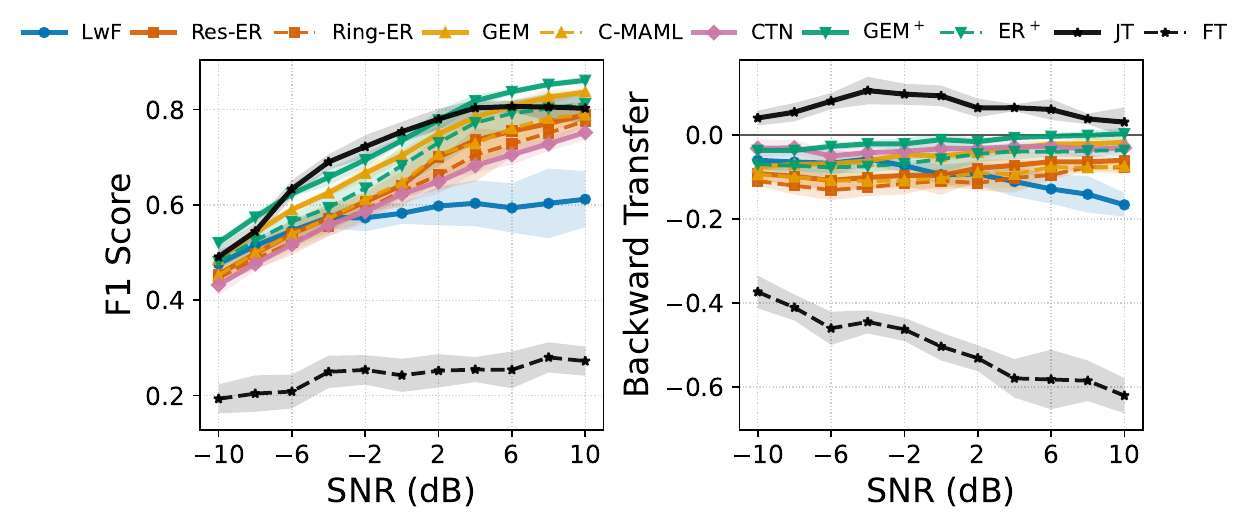}
    \caption{$\fcl$ (left) and \gls{bwt} (right) as a function of SNR for the leading methods in the \gls{1e} \gls{til} setting, with FT and JT for reference. Shading shows the sample standard deviation over training seeds.}
    \label{fig:se_til_snr}
\end{figure}
Fig.~\ref{fig:se_til_snr} shows $\fcl$ as a function of \gls{snr} for the leading methods. All methods cluster at ${-}10$~dB and separate as SNR rises. Notably, GEM overtakes JT at $8{-}10$~dB. The replay methods hold \gls{bwt} roughly constant across SNR, whereas LwF plateaus at ${\sim}60\%$ from ${-}4$~dB with falling \gls{bwt}, as does FT more severely. A larger buffer helps every replay method except A-GEM (Table~\ref{tab:results_mem-sweep_til}). GEM gains the most at every increment and reaches $66.2\%$ at $16384$ samples, closing ${\sim}96\%$ of the FT--JT gap, while CTN and BCL-Dual plateau beyond the default $5120$.

\input{05_Results/Tables/mem-sweep_til}

\subsection{Class-Incremental Setting}
\input{05_Results/Tables/cil_table}
\begin{figure}[t]
    \centering
    \begin{subfigure}{\linewidth}
        \includegraphics[width=\linewidth]{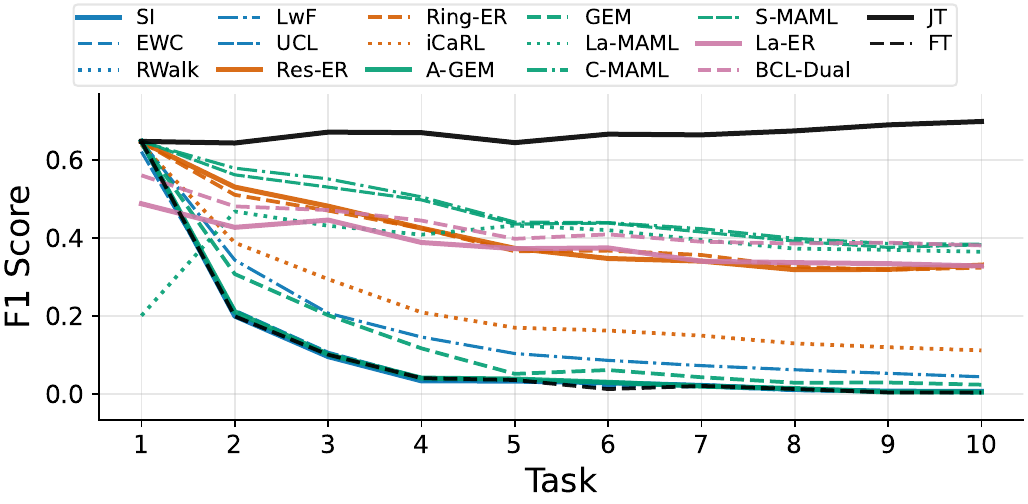}
    \end{subfigure}\\[0em]
    \begin{subfigure}{\linewidth}
        \includegraphics[trim={0cm 0cm 0cm 1.5cm}, clip,width=\linewidth]{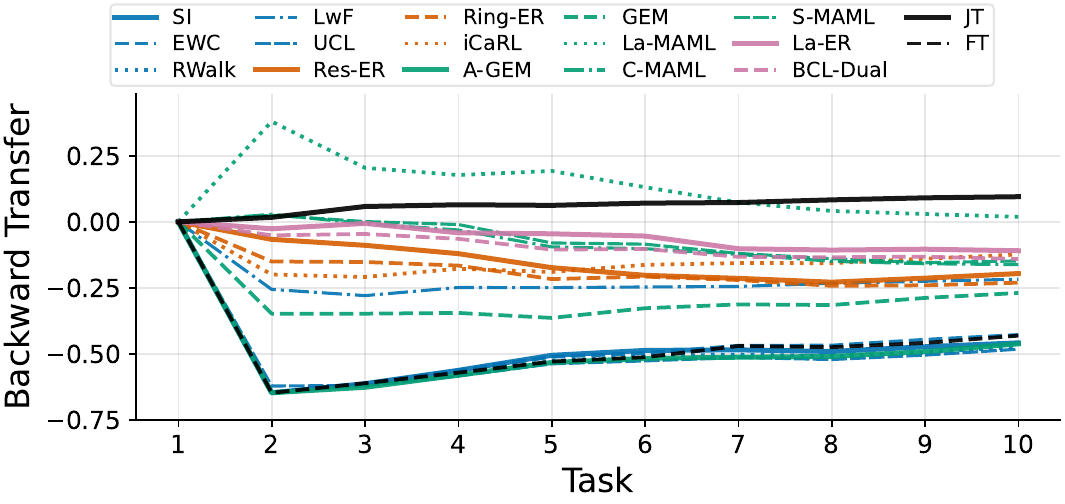}
    \end{subfigure}\\[0em]
    \caption{Per-task \gls{cl} metrics for the \gls{1e} \gls{cil} setting: $\bar{F1}_\text{CL}$ (top) and \gls{bwt} (bottom) as tasks are introduced. Both figures share a legend.}
    \label{fig:se_cil_per_task_metrics}
\end{figure}
\subsubsection{Benchmark}
Table~\ref{tab:results_cil} reports the \num{15} methods that do not require \gls{tid} (CTN, PackNet and HAT are excluded). Removing \gls{tid} costs every method heavily. JT is unaffected ($69.8\%$ against $68.1\%$ under \gls{til}), so a shared head over all classes is not itself harder to learn; the difficulty lies in learning it sequentially. Yet the leading method closes only $55{-}59\%$ of the FT--JT gap. Retention, not acquisition, is the problem: most methods reach ${\sim}65\%$ at Task~1 (Fig.~\ref{fig:se_cil_per_task_metrics}), after which FT and the parameter regularisers collapse to ${\sim}4\%$ by Task~4, while the leaders decline to ${\sim}33{-}38\%$ by Task~10.

The field splits in two. The seven leading \gls{1e} methods lie within $6.0\%$ of one another, and a gap of $21.2\%$ separates them from iCaRL. Every method above the gap trains directly on replayed samples. GEM and A-GEM store samples too, but only to constrain gradients, so the classifier never trains old and new classes together; GEM trails Ring-ER, which has the same buffer, by $30\%$. EWC, SI, RWalk and UCL finish within $0.3\%$ of FT with matching \gls{bwt}: penalising change in important parameters protects task-specific heads, but cannot stop new classes in a shared head from displacing old ones, as observed on image benchmarks~\cite{vandeven2022three}. LwF is the partial exception ($4.5\%$, \gls{bwt} ${-}21.8\%$). Extra epochs lift the leaders by $3.4{-}7.4\%$ and reduce their seed variance, but below the gap only iCaRL and LwF gain more than $2\%$.

\input{05_Results/Tables/cil_ablations_compact}

\subsubsection{Ablating Methods and Combining SS Mechanisms}
Table~\ref{tab:cil_ablations} repeats the ablations on the three replay-based leaders, Res-ER, C-MAML and BCL-Dual. \Gls{1e} \gls{cil} runs are noisy (Res-ER's baseline varies by $\pm6.1$), so several \gls{1e} verdicts are inconclusive; \gls{5e}, with lower variance, is the more decisive. Five findings emerge.
\begin{enumerate}[leftmargin=*]
    \item \textbf{Buffer structure matters beyond utilisation.} Partitioning Res-ER's global buffer by task costs $4.5\%$ in \gls{5e} (E1), and filling the partitioned buffer recovers only half of it (E2, $-2.4$). Unlike \gls{til}, where utilisation explained the gap, a shared head needs every class live in one pool.
    \item \textbf{Meta-learning now matters.} Removing C-MAML's inner-loop adaptation (M2) is \gls{ss} in both regimes ($-6.3\%$/$-1.2\%$), reversing the \gls{til} result. Meta-batch averaging (M1) is \gls{ms} or inconclusive, but is not inherently a meta-learning mechanism.
    \item \textbf{BCL-Dual's bilevel loop is effective, but its validation buffer is harmful.} Collapsing BCL-Dual's bilevel loop is \gls{ss} even at matched budget (B3), so in \gls{cil} its effect is not a compute artefact. Conversely, handing the validation buffer's capacity to replay (B4)\footnote{This reduces BCL-Dual to ``BCL-Single'', a variant proposed by \cite{ll_bcl_2020} but not evaluated here.} \emph{improves} $\fcl$ by $2.3\%$/$3.8\%$. This is consistent with BCL-Dual being limited by its buffer size (Table~\ref{tab:results_mem-sweep_cil}).
    \item \textbf{Distillation improves retention.} Removing \gls{kl} distillation from BCL-Dual costs $1.9\%$/$2.5\%$ (B1, \gls{ms}/\gls{ss}), as in \gls{til}. Adding BCL-Dual's KL distillation to C-MAML (C1) improves $\fcl$ by $+2.6\%$ and $3.6\%$ in \gls{1e} and \gls{5e}, respectively. This helps C-MAML more than BCL-Dual, and the gain is consistent on the held-out seeds. 
    \item \textbf{A global buffer improves BCL-Dual, and the gains are additive.} Pooling BCL-Dual's per-task buffers into one reservoir-sampled buffer (B5) is \gls{ss} ($+1.5\%$/$+2.0\%$), as E1 predicts. Combined with B4, BCL$^+$ (B6) gains $4.3\%$/$6.3\%$, and $3.0\%$/$4.9\%$ on held-out seeds ($41.1\%$/$48.2\%$). This exceeds every method in Table~\ref{tab:results_cil}, and C-MAML$^+$ (C1$^\star$) in \gls{5e} by $3.2\%$, yet JT remains $24{-}29\%$ ahead.
\end{enumerate}

\subsubsection{Robustness to SNR and Buffer Size}
\begin{figure}[t]
    \centering
    \includegraphics[width=\linewidth]{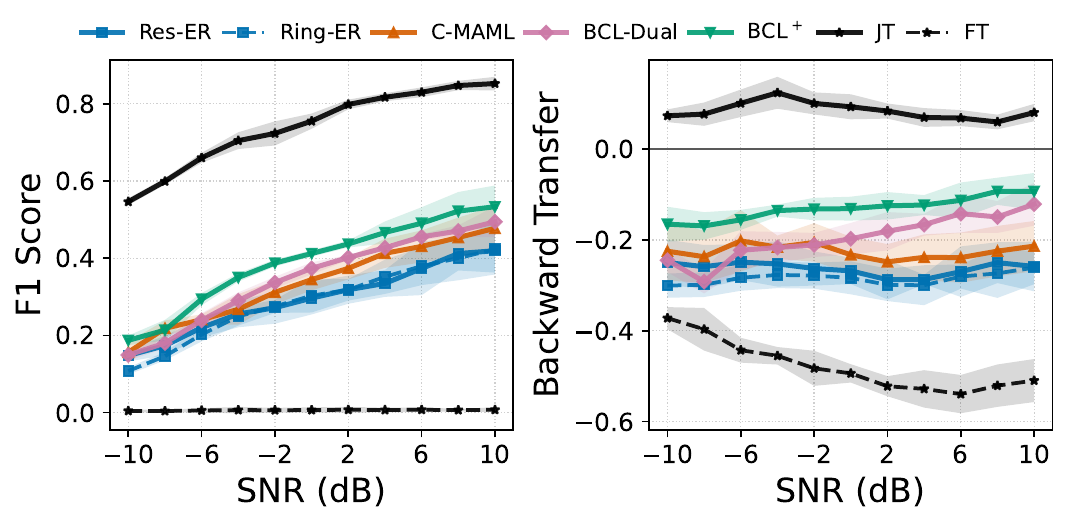}
    \caption{$\fcl$ (left) and \gls{bwt} (right) as a function of SNR for the leading methods in the \gls{1e} \gls{cil} setting, with FT and JT for reference. Shading shows the sample standard deviation over six training seeds.}
    \label{fig:se_cil_snr}
\end{figure}
The leading methods' $\fcl$ rises from ${\sim}11{-}19\%$ at ${-}10$~dB to ${\sim}42{-}54\%$ at ${+}10$~dB (Fig.~\ref{fig:se_cil_snr}), with BCL$^+$ highest at every SNR, but the gap to JT stays at ${\sim}31{-}44\%$ throughout: at ${+}10$~dB the leaders still fall short of JT at ${-}10$~dB. The deficit is therefore not one of noise. \Gls{bwt} is negative for every \gls{cl} method at every SNR; only BCL-Dual's improves with SNR. In the buffer sweep (Table~\ref{tab:results_mem-sweep_cil}), all four methods gain $7.8{-}11.8\%$ up to the default size, beyond which only BCL-Dual improves clearly ($42.0\%$ at $16384$).

\input{05_Results/Tables/mem-sweep_cil}

\subsection{Computational Cost}\label{sec:cost}
\input{05_Results/Tables/mem_footprint.tex}
\begin{figure}[t]
    \centering
    \includegraphics[width=\linewidth]{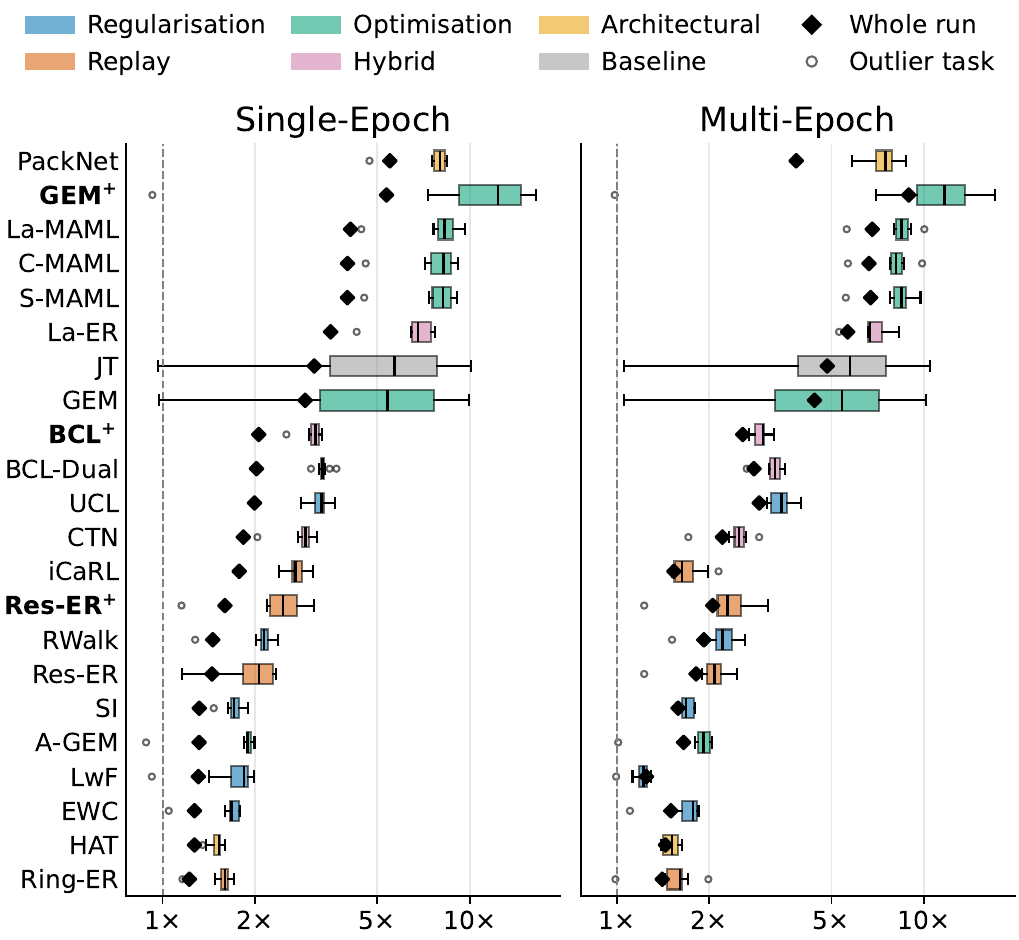}
    \caption{Runtime relative to FT in \gls{1e} and \gls{5e}. Boxes show the per-task distribution; diamonds show the whole run.}
    \label{fig:runtime}
\end{figure}
Table~\ref{tab:storage_breakdown_til} breaks down each method's memory footprint. GEM is the heaviest at $182.8$~MB, over $6{\times}$ the lightest, because it stores one gradient per previous task ($10\,|\theta|$ at $T{=}10$); A-GEM avoids this growth by keeping a single averaged gradient. The parameter regularisers differ only in the number of per-parameter buffers (three for EWC, four for SI, six for RWalk), with no return in $\fcl$, and the adaptive-rate MAML variants each carry an extra $|\theta|$ yet trail C-MAML. Replay methods cost ${\sim}35$~MB, and LwF is the lightest at $29.6$~MB.

Runtime (Fig.~\ref{fig:runtime}) shows a similar trend. Most replay and regularisation methods run within $1.2{-}2\times$ FT, whereas the MAML variants cost ${\sim}4\times$ in \gls{1e} and ${\sim}6{-}7\times$ in \gls{5e}. PackNet is the slowest in \gls{1e} due to needing post-pruning fine-tuning. GEM's per-task cost grows with the number of stored constraints, spanning $1{-}10\times$ across tasks. Read against the ablations, the meta-learning overhead buys nothing in \gls{til}, where Res-ER matches C-MAML at ${\sim}1.5\times$, but is justified in \gls{cil}, where inner-loop adaptation is statistically significant. The combined models (marked ${^+}$) contribute only stored soft targets to each footprint, so each combined model costs about the same as the method it is built on. BCL$^+$ therefore gives the best \gls{cil} result, and Res-ER$^+$ comes within $2.3\%$ of GEM$^+$ in \gls{5e}, both at replay-level cost (${\sim}35$~MB, ${\sim}1.5{-}2\times$ FT), against $183$~MB for GEM$^+$.

%% file: 05_Results/Tables/til_table.tex
\begin{table}[t]
\centering
\caption{Continual learning performance for the TIL setting under \gls{1e} and \gls{5e} training. Results show the mean and sample standard deviation (in $\%$) over training seeds. $\Delta\fcl$ is the change in $\fcl$ from \acrshort{1e} to \acrshort{5e}.}
\label{tab:results_til}
\setlength{\tabcolsep}{4pt}
\resizebox{\columnwidth}{!}{%
\begin{tabular}{l|c||c|c|c||c|c|c||c}
\hline\toprule
 & & \multicolumn{3}{c||}{\textbf{\acrshort{1e}}} & \multicolumn{3}{c||}{\textbf{\acrshort{5e}}} & \\
\cmidrule(lr){3-5} \cmidrule(lr){6-8}
\textbf{Method} & \textbf{Approach} & $\mathbf{\fcl}$ & \textbf{BWT} & \textbf{FWT} & $\mathbf{\fcl}$ & \textbf{BWT} & \textbf{FWT} & $\mathbf{\Delta\fcl}$ \\
\hline\toprule
FT & N/A & \pmv{21.2}{2.4} & \pmv{-47.3}{2.8} & \pmv{1.9}{1.1} & \pmv{25.3}{1.1} & \pmv{-60.9}{1.5} & \pmv{1.7}{1.3} & $+4.1$ \\
\midrule
A-GEM & Optim. & \pmv{34.6}{3.4} & \pmv{-29.3}{2.1} & \pmv{2.5}{1.3} & \pmv{35.5}{2.8} & \pmv{-50.0}{3.2} & \pmv{2.0}{1.3} & $+0.9$ \\
UCL & Reg. & \pmv{37.3}{4.8} & \pmv{-31.3}{4.9} & \pmv{1.0}{1.6} & \pmv{37.2}{4.2} & \pmv{-44.3}{5.1} & \pmv{0.2}{1.3} & $-0.1$ \\
iCaRL & Replay & \pmv{42.9}{4.5} & \pmv{-2.5}{3.0} & \pmv{0.5}{1.2} & \pmv{61.1}{3.8} & \pmv{-5.5}{2.3} & \pmv{-0.5}{1.6} & $+18.2$ \\
La-ER & Hybrid & \pmv{50.3}{2.1} & \pmv{-5.6}{1.7} & \pmv{1.0}{0.4} & \pmv{58.2}{0.5} & \pmv{-11.4}{0.9} & \pmv{1.2}{0.6} & $+7.9$ \\
HAT & Arch. & \pmv{50.3}{1.9} & \pmv{-8.7}{1.8} & \pmv{1.5}{0.7} & \pmv{60.6}{1.1} & \pmv{-6.6}{1.0} & \pmv{1.2}{1.0} & $+10.3$ \\
RWalk & Reg. & \pmv{51.1}{3.7} & \pmv{-5.7}{3.1} & \pmv{1.5}{1.2} & \pmv{57.9}{2.7} & \pmvdul{-0.3}{0.4} & \pmv{1.6}{1.2} & $+6.8$ \\
SI & Reg. & \pmv{51.3}{3.0} & \pmvdul{-0.8}{0.8} & \pmv{1.6}{1.1} & \pmv{57.3}{2.1} & \pmvbf{0.2}{0.1} & \pmv{1.5}{1.0} & $+6.0$ \\
EWC & Reg. & \pmv{51.3}{3.3} & \pmv{-3.2}{2.0} & \pmv{1.4}{1.2} & \pmv{59.3}{2.4} & \pmv{-1.2}{0.8} & \pmv{0.8}{1.0} & $+8.0$ \\
PackNet & Arch. & \pmv{51.4}{2.4} & \pmvul{0.1}{0.1} & \pmvbf{3.9}{2.1} & \pmv{58.9}{2.1} & \pmvbf{0.2}{0.1} & \pmvbf{3.8}{1.2} & $+7.5$ \\
LwF & Reg. & \pmv{52.8}{3.3} & \pmv{-9.8}{3.0} & \pmvul{3.4}{1.3} & \pmvdul{67.0}{2.3} & \pmv{-11.1}{1.5} & \pmvul{3.0}{0.9} & $+14.2$ \\
La-MAML & Optim. & \pmv{54.7}{1.9} & \pmvbf{19.6}{4.8} & \pmv{1.4}{1.4} & \pmv{63.9}{1.2} & \pmv{-3.0}{6.0} & \pmv{2.1}{1.3} & $+9.2$ \\
BCL-Dual & Hybrid & \pmv{56.6}{2.0} & \pmv{-8.1}{0.8} & \pmvdul{2.7}{0.6} & \pmv{64.6}{2.7} & \pmv{-13.3}{1.4} & \pmvdul{2.9}{1.3} & $+8.0$ \\
CTN & Hybrid & \pmv{56.6}{0.7} & \pmv{-3.0}{0.7} & \pmv{1.5}{0.9} & \pmv{62.7}{1.2} & \pmv{-10.0}{0.7} & \pmv{1.3}{1.2} & $+6.1$ \\
Ring-ER & Replay & \pmv{57.1}{1.3} & \pmv{-10.5}{1.7} & \pmv{2.0}{1.5} & \pmv{65.7}{0.9} & \pmv{-15.5}{0.5} & \pmv{1.8}{1.0} & $+8.6$ \\
Res-ER & Replay & \pmv{59.1}{1.4} & \pmv{-7.8}{1.3} & \pmv{1.6}{1.0} & \pmvul{67.3}{1.2} & \pmv{-13.2}{1.1} & \pmv{1.9}{1.0} & $+8.2$ \\
C-MAML & Optim. & \pmvdul{59.4}{1.0} & \pmv{-8.4}{1.5} & \pmv{1.9}{0.8} & \pmvdul{67.0}{0.7} & \pmv{-13.5}{0.5} & \pmv{1.6}{1.2} & $+7.6$ \\
S-MAML & Optim. & \pmvul{60.0}{1.5} & \pmv{-7.0}{1.0} & \pmv{1.6}{1.0} & \pmv{66.1}{0.6} & \pmv{-12.9}{0.7} & \pmv{1.8}{1.4} & $+6.1$ \\
GEM & Optim. & \pmvbf{62.8}{1.5} & \pmv{-6.4}{1.2} & \pmv{2.3}{0.9} & \pmvbf{68.7}{0.6} & \pmv{-12.8}{0.5} & \pmv{1.8}{1.1} & $+5.9$ \\
\midrule
JT & N/A & \pmv{68.1}{1.8} & \pmv{9.2}{1.9} & \pmv{1.9}{0.6} & \pmv{73.0}{1.2} & \pmv{1.8}{0.9} & \pmv{1.8}{1.0} & $+4.9$ \\
\bottomrule\hline
\end{tabular}%
}
\end{table}

%% file: 05_Results/Tables/til_ablations_compact.tex
\begin{table}[t]
\centering
\caption{\gls{til} ablations. Each row removes or replaces ($\to$) one mechanism relative to its full method (ID~0). $\fcl$ is the mean over $6$ or $^\dagger 12$ seeds; verdicts (\textbf{V.}) follow Section~\ref{sec:stat_protocol}. ${^+}$denotes a model combining SS mechanisms; $^\star$means tested on held-out seeds (i.e. same seeds as Table~\ref{tab:results_til}). Full results are in Appendix~\ref{apx:full-ablations}.}
\label{tab:til_ablations}
\setlength{\tabcolsep}{4pt}
\resizebox{\columnwidth}{!}{%
\begin{tabular}{llcclccl}
\hline\toprule
 & & \multicolumn{3}{c}{\textbf{\acrshort{1e}}} & \multicolumn{3}{c}{\textbf{\acrshort{5e}}} \\
\cmidrule(lr){3-5} \cmidrule(lr){6-8}
\textbf{ID} & \textbf{Mechanism/Description} & $\mathbf{\fcl}$ & $\mathbf{\Delta}\fcl$ & \textbf{V.} & $\mathbf{\fcl}$ & $\mathbf{\Delta}\fcl$ & \textbf{V.} \\
\hline\toprule
G0 & GEM & $62.83$ & -- & & $68.54$ & -- & \\
G1 & Projection ($\to$ ring replay) & $58.07$ & $-4.76$ & \textbf{SS} & $65.39$ & $-3.16$ & \textbf{SS} \\
\midrule
E0 & Res-ER & $58.89^\dagger$ & -- & & $66.94$ & -- & \\
E1 & Reservoir ($\to$ static ring) & $57.44^\dagger$ & $-1.45$ & \textbf{SS} & $65.39$ & $-1.56$ & \textbf{SS} \\
E2 & Reservoir ($\to$ full ring) & $58.88$ & $-0.08$ & NS & $66.40$ & $-0.54$ & \textbf{SS} \\
\midrule
M0 & C-MAML & $59.71^\dagger$ & -- & & $63.91^\dagger$ & -- & \\
M1 & Meta-batch averaging & $60.56$ & $+0.62$ & MS & $64.66^\dagger$ & $+0.75$ & \textbf{SS} \\
M2 & Inner-loop adaptation & $59.65^\dagger$ & $-0.06$ & NS & $65.63$ & $+1.38$ & \textbf{SS} \\
\midrule
B0 & BCL-Dual & $56.88^\dagger$ & -- & & $65.37^\dagger$ & -- & \\
B1 & \gls{kl} distillation & $55.14$ & $-1.52$ & \textbf{SS} & $62.73$ & $-2.96$ & \textbf{SS} \\
B2 & Bilevel interpolation & $55.26$ & $-1.40$ & MS & $63.35^\dagger$ & $-2.03$ & \textbf{SS} \\
B3 & Bilevel loop (matched budget) & $55.37^\dagger$ & $-1.51$ & MS & $65.59^\dagger$ & $+0.22$ & NS \\
B4 & Validation buffer (outer step) & $54.77^\dagger$ & $-2.11$ & \textbf{SS} & $65.23^\dagger$ & $-0.14$ & INC \\
\midrule
T0 & CTN & $57.17^\dagger$ & -- & & $61.63^\dagger$ & -- & \\
T1 & FiLM conditioning & $56.55^\dagger$ & $-0.62$ & \textbf{SS} & $61.35^\dagger$ & $-0.27$ & NS \\
T2 & \gls{kl} distillation & $56.15$ & $-1.17$ & \textbf{SS} & $59.21$ & $-2.88$ & \textbf{SS} \\
\hline\toprule
\multicolumn{8}{l}{\textit{GEM$^+$ (A1--2 is tested vs.\ A0; C1 vs.\ A1)}} \\
A0 & GEM & $62.83$ & -- & & $68.54$ & -- & \\
A1 & A0 $+$ Dynamic ring & $63.95$ & $+1.12$ & \textbf{SS} & $69.19$ & $+0.65$ & \textbf{SS} \\
A2 & A0 $+$ \gls{kl} distillation & $65.60$ & $+2.77$ & \textbf{SS} & $72.75$ & $+4.21$ & \textbf{SS} \\
C1 & A1 $+$ \gls{kl} (tested vs. A1) & $66.51$ & $+2.56$ & \textbf{SS} & $73.28$ & $+4.09$ & \textbf{SS} \\
C1$^\star$ & C1 tested on held-out seeds & $66.17$ & $+3.39$ & \textbf{SS} & $74.22$ & $+5.52$ & \textbf{SS} \\
\midrule
\multicolumn{8}{l}{\textit{Res-ER$^+$ (R1 vs.\ E0)}} \\
E0 & Res-ER & $58.89^\dagger$ & -- & & $66.94$ & -- & \\
R1 & E0 $+$ \gls{kl} distillation & $62.10$ & $+3.15$ & \textbf{SS} & $71.69$ & $+4.75$ & \textbf{SS} \\
R1$^\star$ & R1 tested on held-out seeds & $61.70$ & $+2.55$ & \textbf{SS} & $71.91$ & $+4.61$ & \textbf{SS} \\
\bottomrule\hline
\end{tabular}%
}
\end{table}

%% file: 05_Results/Tables/mem-sweep_til.tex
\begin{table}[t]
\centering
\caption{Memory buffer size vs $\bar{F1}_\text{CL}$ for methods utilising a memory buffer, in the \gls{1e} \gls{til} setting.}
\label{tab:results_mem-sweep_til}
\resizebox{\columnwidth}{!}{%
\begin{tabular}{lccccc}
\hline\toprule
\textbf{Method} & \textbf{1024} & \textbf{2048} & \textbf{5120} & \textbf{8192} & \textbf{16384} \\
\hline\toprule
A-GEM & \pmv{37.0}{2.4} & \pmv{36.2}{2.8} & \pmv{34.6}{3.4} & \pmv{34.9}{2.3} & \pmv{36.5}{3.0} \\
CTN & \pmv{54.0}{0.8} & \pmv{55.0}{0.7} & \pmv{56.6}{0.7} & \pmv{56.8}{1.1} & \pmv{56.8}{1.5} \\
BCL-Dual & \pmv{52.3}{1.5} & \pmv{54.2}{2.5} & \pmv{56.6}{2.0} & \pmv{57.1}{2.5} & \pmv{57.5}{2.5} \\
Ring-ER & \pmv{52.5}{1.5} & \pmv{54.8}{1.6} & \pmv{57.1}{1.3} & \pmv{58.2}{1.2} & \pmv{59.3}{1.5} \\
Res-ER & \pmv{54.6}{1.1} & \pmv{56.4}{1.6} & \pmv{59.1}{1.4} & \pmv{60.3}{1.7} & \pmv{59.6}{3.2} \\
C-MAML & \pmv{54.4}{2.5} & \pmv{57.1}{1.4} & \pmv{59.4}{1.0} & \pmv{60.6}{1.3} & \pmv{61.4}{1.5} \\
GEM & \pmv{57.7}{1.0} & \pmv{60.4}{1.2} & \pmv{62.8}{1.5} & \pmv{64.6}{1.3} & \pmv{66.2}{1.6} \\
\midrule
Res-ER$^+$ & \pmv{58.7}{1.8} & \pmv{60.9}{2.1} & \pmv{61.7}{1.9}& \pmv{62.3}{1.9} & \pmv{62.2}{2.1} \\
GEM$^+$ & \pmvbf{62.0}{1.1} & \pmvbf{63.8}{1.8} & \pmvbf{66.2}{1.7} & \pmvbf{67.2}{1.6} & \pmvbf{68.5}{2.1} \\
\bottomrule\hline
\end{tabular}%
}
\end{table}

%% file: 05_Results/Tables/cil_table.tex
\begin{table}[t]
\centering
\caption{Continual learning performance for the CIL setting under \gls{1e} and \gls{5e} training. Results show the mean and sample standard deviation (in $\%$) over training seeds. $\Delta\fcl$ is the change in $\fcl$ from \acrshort{1e} to \acrshort{5e}.}
\label{tab:results_cil}
\setlength{\tabcolsep}{4pt}
\resizebox{\columnwidth}{!}{%
\begin{tabular}{l|c||c|c|c||c|c|c||c}
\hline\toprule
 & & \multicolumn{3}{c||}{\textbf{\acrshort{1e}}} & \multicolumn{3}{c||}{\textbf{\acrshort{5e}}} & \\
\cmidrule(lr){3-5} \cmidrule(lr){6-8}
\textbf{Method} & \textbf{Approach} & $\mathbf{\fcl}$ & \textbf{BWT} & \textbf{FWT} & $\mathbf{\fcl}$ & \textbf{BWT} & \textbf{FWT} & $\mathbf{\Delta\fcl}$ \\
\hline\toprule
FT & N/A & \pmv{0.4}{0.4} & \pmv{-43.1}{2.8} & \pmv{-1.1}{0.3} & \pmv{0.9}{0.5} & \pmv{-55.3}{3.1} & \pmv{-1.2}{0.3} & $+0.5$\\
\midrule
EWC & Reg. & \pmv{0.4}{0.3} & \pmv{-42.9}{5.4} & \pmv{-1.1}{0.3} & \pmv{0.9}{0.5} & \pmv{-56.1}{3.1} & \pmv{-1.2}{0.3} & $+0.5$\\
RWalk & Reg. & \pmv{0.5}{0.5} & \pmv{-47.4}{2.6} & \pmv{-1.1}{0.3} & \pmv{1.2}{0.4} & \pmv{-64.1}{2.9} & \pmv{-1.2}{0.3} & $+0.7$\\
A-GEM & Optim. & \pmv{0.6}{0.5} & \pmv{-46.3}{2.1} & \pmv{-1.1}{0.3} & \pmv{0.9}{0.3} & \pmv{-56.0}{3.0} & \pmv{-1.2}{0.3} & $+0.3$\\
UCL & Reg. & \pmv{0.6}{0.5} & \pmv{-48.3}{3.7} & \pmv{-1.0}{0.4} & \pmv{1.2}{0.5} & \pmv{-65.5}{3.7} & \pmv{-1.0}{0.4} & $+0.6$\\
SI & Reg. & \pmv{0.7}{0.4} & \pmv{-45.9}{5.2} & \pmv{-1.1}{0.3} & \pmv{1.0}{0.4} & \pmv{-59.4}{3.3} & \pmv{-1.2}{0.3} & $+0.3$\\
GEM & Optim. & \pmv{2.4}{1.2} & \pmv{-27.0}{4.6} & \pmv{-1.1}{0.3} & \pmv{4.2}{1.0} & \pmv{-48.5}{5.5} & \pmv{-1.2}{0.3} & $+1.8$\\
LwF & Reg. & \pmv{4.5}{0.7} & \pmv{-21.8}{3.5} & \pmv{-1.2}{0.3} & \pmv{8.3}{1.9} & \pmv{-29.5}{2.7} & \pmv{-1.2}{0.3} & $+3.8$\\
iCaRL & Replay & \pmv{11.2}{1.8} & \pmvdul{-12.4}{6.2} & \pmv{-1.2}{0.3} & \pmv{28.8}{3.3} & \pmvbf{-13.8}{2.2} & \pmv{-1.2}{0.3} & $+17.6$\\
Ring-ER & Replay & \pmv{32.4}{1.1} & \pmv{-23.1}{2.4} & \pmv{-1.1}{0.3} & \pmv{37.2}{1.0} & \pmv{-33.8}{2.1} & \pmv{-1.2}{0.3} & $+4.8$\\
La-ER & Hybrid & \pmv{32.8}{2.0} & \pmvul{-11.0}{3.9} & \pmv{-0.8}{0.6} & \pmv{36.2}{1.8} & \pmvul{-20.9}{2.9} & \pmv{-0.8}{0.6} & $+3.4$\\
Res-ER & Replay & \pmv{33.1}{3.2} & \pmv{-19.6}{3.3} & \pmv{-1.1}{0.3} & \pmv{40.5}{1.4} & \pmv{-30.8}{2.7} & \pmv{-1.2}{0.3} & $+7.4$\\
La-MAML & Optim. & \pmv{36.4}{2.5} & \pmvbf{1.9}{7.0} & \pmv{-1.2}{0.3} & \pmv{41.8}{1.1} & \pmvdul{-21.1}{2.9} & \pmv{-1.2}{0.3} & $+5.4$\\
BCL-Dual & Hybrid & \pmvdul{38.1}{2.1} & \pmv{-14.1}{4.1} & \pmv{-1.2}{0.3} & \pmvbf{43.3}{1.0} & \pmv{-23.8}{2.2} & \pmv{-1.2}{0.3} & $+5.2$\\
S-MAML & Optim. & \pmvul{38.3}{3.1} & \pmv{-14.8}{3.8} & \pmv{-1.2}{0.3} & \pmvul{42.7}{0.9} & \pmv{-28.3}{2.3} & \pmv{-1.2}{0.3} & $+4.4$\\
C-MAML & Optim. & \pmvbf{38.4}{1.6} & \pmv{-16.2}{2.8} & \pmv{-1.1}{0.3} & \pmvdul{42.4}{1.1} & \pmv{-29.1}{1.9} & \pmv{-1.2}{0.3} & $+4.0$\\
\midrule
JT & N/A & \pmv{69.8}{2.2} & \pmv{9.5}{2.2} & \pmv{-1.1}{0.3} & \pmv{72.5}{3.3} & \pmv{0.1}{2.0} & \pmv{-1.2}{0.3} & $+2.7$\\
\bottomrule\hline
\end{tabular}%
}
\end{table}

%% file: 05_Results/Tables/cil_ablations_compact.tex
\begin{table}[t]
\centering
\caption{\gls{cil} ablations. Each row removes or replaces ($\to$) one mechanism relative to its full method (ID~0). $\fcl$ is the mean over $6$ or $^\dagger 12$ seeds; verdicts (\textbf{V.}) follow Section~\ref{sec:stat_protocol}. ${^+}$denotes a model combining SS mechanisms; $^\star$means tested on held-out seeds (i.e. same seeds as Table~\ref{tab:results_cil}). Full results are in Appendix~\ref{apx:full-ablations}.}
\label{tab:cil_ablations}
\setlength{\tabcolsep}{4pt}
\resizebox{\columnwidth}{!}{%
\begin{tabular}{llcclccl}
\hline\toprule
 & & \multicolumn{3}{c}{\textbf{\acrshort{1e}}} & \multicolumn{3}{c}{\textbf{\acrshort{5e}}} \\
\cmidrule(lr){3-5} \cmidrule(lr){6-8}
\textbf{ID} & \textbf{Mechanism/Description} & $\mathbf{\fcl}$ & $\mathbf{\Delta}\fcl$ & \textbf{V.} & $\mathbf{\fcl}$ & $\mathbf{\Delta}\fcl$ & \textbf{V.} \\
\hline\toprule
E0 & Res-ER & $31.13^\dagger$ & -- & & $40.50^\dagger$ & -- & \\
E1 & Reservoir ($\to$ static ring) & $32.08^\dagger$ & $+0.95$ & INC & $36.37$ & $-4.45$ & \textbf{SS} \\
E2 & Reservoir ($\to$ full ring) & $31.94^\dagger$ & $+0.81$ & INC & $38.47$ & $-2.36$ & \textbf{SS} \\
\midrule
M0 & C-MAML & $37.17^\dagger$ & -- & & $41.86^\dagger$ & -- & \\
M1 & Meta-batch averaging & $35.06$ & $-3.72$ & MS & $41.38^\dagger$ & $-0.48$ & INC \\
M2 & Inner-loop adaptation & $32.53$ & $-6.25$ & \textbf{SS} & $40.65^\dagger$ & $-1.22$ & \textbf{SS} \\
\midrule
B0 & BCL-Dual & $37.47^\dagger$ & -- & & $42.26^\dagger$ & -- & \\
B1 & \gls{kl} distillation & $36.43$ & $-1.88$ & MS & $39.92$ & $-2.54$ & \textbf{SS} \\
B2 & Bilevel interpolation & $37.42^\dagger$ & $-0.05$ & INC & $42.77^\dagger$ & $+0.51$ & INC \\
B3 & Bilevel loop (matched budget) & $31.50$ & $-6.81$ & \textbf{SS} & $39.58$ & $-2.87$ & \textbf{SS} \\
B4 & No validation buffer & $40.61$ & $+2.30$ & \textbf{SS} & $46.30$ & $+3.84$ & \textbf{SS} \\
\hline\toprule
\multicolumn{8}{l}{\textit{C-MAML$^+$ (C1 vs.\ M0 on 6 seeds)}} \\
M0 & C-MAML & $38.78$ & -- & & $41.58$ & -- & \\
C1 & M0 $+$ \gls{kl} (slow weights) & $41.39$ & $+2.60$ & \textbf{SS} & $45.18$ & $+3.61$ & \textbf{SS} \\
C1$^\star$ & C1 tested on held-out seeds & $41.09$ & $+2.72$ & \textbf{SS} & $45.04$ & $+2.67$ & \textbf{SS} \\
\midrule
\multicolumn{8}{l}{\textit{BCL$^+$ (= B6)}} \\
B4$^\star$ & B4 tested on held-out seeds & $40.35$ & $+2.27$ & MS & $46.08$ & $+2.74$ & \textbf{SS} \\
B5 & Global reservoir & $39.82$ & $+1.51$ & \textbf{SS} & $44.41$ & $+1.96$ & \textbf{SS} \\
B6 & B4 $+$ global reservoir & $42.64$ & $+4.33$ & \textbf{SS} & $48.79$ & $+6.34$ & \textbf{SS} \\
B6$^\star$ & B6 tested on held-out seeds & $41.12$ & $+3.04$ & \textbf{SS} & $48.23$ & $+4.89$ & \textbf{SS} \\
\bottomrule\hline
\end{tabular}%
}
\end{table}

%% file: 05_Results/Tables/mem-sweep_cil.tex
\begin{table}[t]
\centering
\caption{Memory buffer size vs $\bar{F1}_\text{CL}$ for methods utilising a memory buffer, in the \gls{1e} \gls{cil} setting. BCL$^+$ replaces BCL-Dual's per-task buffers with one reservoir-sampled buffer and hands its validation buffer to replay.}
\label{tab:results_mem-sweep_cil}
\resizebox{\columnwidth}{!}{%
\begin{tabular}{lccccc}
\hline\toprule
\textbf{Method} & \textbf{1024} & \textbf{2048} & \textbf{5120} & \textbf{8192} & \textbf{16384} \\
\hline\toprule
Res-ER & \pmv{25.3}{1.7} & \pmv{31.0}{1.6} & \pmv{33.1}{3.2} & \pmv{30.3}{5.0} & \pmv{31.1}{4.7} \\
Ring-ER & \pmv{20.6}{2.9} & \pmv{26.6}{2.9} & \pmv{32.4}{1.1} & \pmv{33.2}{2.2} & \pmv{34.0}{2.9} \\
C-MAML & \pmv{27.1}{2.2} & \pmv{32.9}{1.4} & \pmv{38.4}{1.6} & \pmv{39.2}{2.0} & \pmv{37.8}{4.1} \\
BCL-Dual & \pmv{27.9}{1.8} & \pmv{33.3}{1.4} & \pmv{38.1}{2.1} & \pmv{40.0}{2.1} & \pmv{42.0}{1.7} \\
\midrule
BCL$^+$ & \pmvbf{34.0}{1.2} & \pmvbf{37.5}{1.4} & \pmvbf{41.1}{1.4} & \pmvbf{43.6}{2.0} & \pmvbf{43.0}{2.4} \\
\bottomrule\hline
\end{tabular}%
}
\end{table}

%% file: 05_Results/Tables/mem_footprint.tex
\begin{table}[t]
\centering
\caption{Memory footprint breakdown per \gls{cl} method in MB, where the base model size $\lvert\theta\rvert {=} 14.80$~MB. Columns denote memory consumed by the model parameters, replay buffer, regularisation terms, and architectural components, respectively.}
\label{tab:storage_breakdown_til}
\footnotesize
\resizebox{0.8\columnwidth}{!}{%
\begin{tabular}{lccccc}
\hline\toprule
\textbf{Method} & \textbf{Model} & \textbf{Replay} & \textbf{Reg} & \textbf{Arch} & \textbf{Total} \\
\hline\toprule
\textbf{GEM$^+$}       & $|\theta|$       & 20.2   & $10\,|\theta|$  & 0     & 182.9 \\
GEM        & $|\theta|$       & 20.1   & $10\,|\theta|$  & 0                   & 182.8          \\
RWalk      & $|\theta|$       & 0      & $6\,|\theta|$   & 0                   & 103.6          \\
SI         & $|\theta|$       & 0      & $4\,|\theta|$   & 0                   & 74.0           \\
EWC        & $|\theta|$       & 0      & $3\,|\theta|$   & 0                   & 59.2           \\
UCL        & $2\,|\theta|$    & 0      & $2\,|\theta|$   & 0                   & 59.0           \\
La-MAML    & $2\,|\theta|$    & 20.5   & 0               & 0                   & 50.1           \\
S-MAML     & $2\,|\theta|$    & 20.5   & 0               & 0                   & 50.1           \\
La-ER      & $2\,|\theta|$    & 20.1   & 0               & 0                   & 49.7           \\
iCaRL      & $|\theta|$       & 20.6   & 0               & 0                   & 35.4           \\
C-MAML     & $|\theta|$       & 20.5   & 0               & 0                   & 35.3           \\
CTN        & $1.02\,|\theta|$ & 20.2   & 0               & 0                   & 35.2           \\
\textbf{BCL$^+$}  & $|\theta|$       & 20.2   & 0               & 0     & 35.0  \\
\textbf{Res-ER$^+$}        & $|\theta|$       & 20.2   & 0               & 0     & 35.0 \\
BCL-Dual   & $|\theta|$       & 20.2   & 0               & 0                   & 35.0           \\
Ring-ER    & $|\theta|$       & 20.2   & 0               & 0                   & 35.0           \\
A-GEM      & $|\theta|$       & 20.1   & 0               & 0                   & 34.9           \\
Res-ER     & $|\theta|$       & 20.1   & 0               & 0                   & 34.9           \\
PackNet    & $|\theta|$       & 0      & 0               & $1.25\,|\theta|$    & 33.3           \\
HAT        & $1.01\,|\theta|$ & 0      & 0               & $|\theta|$          & 29.8           \\
LwF        & $|\theta|$       & 0      & $|\theta|$      & 0                   & \textbf{29.6}  \\
\bottomrule\hline
\end{tabular}%
}
\end{table}

%% file: Conclusion.tex
\glsresetall
\section{CONCLUSION}
\label{sec:conclusion}

Radar waveforms proliferate faster than a classifier can be rebuilt around them. Parameter-agile and adaptive waveforms are widening the classification problem faster than fixed-class training can follow. Retraining from scratch means revisiting every waveform ever collected, with a linear cost. \gls{cl} enables classifiers to learn new waveforms sequentially, whilst retaining prior knowledge and without retraining from scratch. Its methods, however, are mostly validated on image datasets. Prior studies on radar datasets are limited to only a handful of \gls{cl} methods evaluated on few classes. 

This paper presents the first large-scale benchmark of \gls{cl} methods for radar waveform classification. Eighteen methods are benchmarked on a ten-task problem spanning \num{54} waveforms, evaluated under \gls{til} and \gls{cil} settings, and in both \gls{1e} and \gls{5e} regimes. Performance is attributed to mechanisms rather than to whole methods via leave-one-out ablations. Ablations are evaluated using statistical significance testing. Statistically significant mechanisms are then combined, producing the strongest methods in all settings.

Methods that retain data, either for replay or gradient projection, lead the benchmark. Under \gls{til}, GEM leads both regimes, and the ablations attribute its lead to gradient projection. Distillation (BCL-Dual, CTN) and full buffer utilisation (Res-ER) are also \textit{statistically significant}, whereas BCL-Dual's bilevel components and CTN's FiLM conditioning are significant only in \gls{1e}. Meta-learning is either \textit{not significant} or is actively harmful under TIL. Combining SS mechanisms (gradient projection, dynamic ring buffering and \gls{kl} distillation) produces the best TIL model. A lightweight alternative, Res-ER$^+$, beats all others except GEM in \gls{1e}.

The \gls{cil} setting exposes a harder problem. JT is unaffected by the loss of \gls{tid}, yet the leading CIL results compare to the worst TIL results. The key challenge lies in the shared classifier head, which must discriminate a growing label space. Only methods that train directly on replayed samples are relatively effective. Parameter regularisers are within $0.3\%$ of FT; and GEM, which uses its buffer only to constrain gradients, trails Ring-ER by $30\%$. Likewise, ablation results find that buffer structure matters beyond utilisation. Meta-learning, in contrast to \gls{til}, becomes significant, especially for \gls{1e}. BCL-Dual's bilevel loop is also significant, but its validation buffer is harmful. Like for TIL, the combined model, BCL$^+$ ($41.1\%$/$48.2\%$), exceeds every other method.

These results give developers of adaptive radar classification systems a mechanism-level basis for design. Where past data may be stored, replay is the critical component, and it should be paired with distillation. Distillation improves retention at little added memory and compute costs. Under \gls{til}, gradient projection has the highest $\fcl$ at roughly $5{\times}$ the memory and $4{\times}$ the compute of replay. Under \gls{cil}, the buffer should be global, and BCL$^+$ gives the best result at replay-level cost. Where past data may not be stored, distilling against a frozen teacher (LwF~\cite{ll_lwf_2017}) is the strongest alternative, although under \gls{cil} it reaches only $8.3\%$.

This work makes some simplifying assumptions. Task boundaries are given, every new sample is correctly labelled, and the system is told when a waveform is novel. In practice, novelty must first be detected, which is an open-set recognition problem~\cite{2013_scheirer_osr}, and labels may be noisy or absent. Relaxing these assumptions, through task-free~\cite{2019_aljundi_taskfree} and open-set \gls{cl}, would bring the models closer to deployment.

Two further directions follow from this analysis. The first is generative replay, which preserves the replay mechanism using synthetic data from a generative model trained on past data. The second is task inference at test time. The gap between the two settings, $59.1\%$ against $33.1\%$ $\fcl$ for \gls{1e} Res-ER, is the value of knowing \gls{tid} at inference. Predicting it from the signal, rather than requiring it, may recover part of that difference.

%% file: 07_Appendix/Appendix.tex
\appendix
\input{07_Appendix/overlapping_mods}
\input{07_Appendix/tid_leakage.tex}
\input{07_Appendix/hpt.tex}
\input{07_Appendix/full_ablations}

%% file: 07_Appendix/overlapping_mods.tex
\subsection{Cross-Dataset Class Overlap}
\label{apx:overlapping-mods}

The four source datasets were generated independently, and several waveform classes use the same modulation. An overlap is only counted when the specific waveform matches (e.g. Barker-5 with Barker-5, not Barker with Barker). This appendix empirically shows that classes with the same specific modulation are distinguishable from each other. Table~\ref{tab:overlap-pairs} lists the $13$ resulting pairs. Barker and Frank codes occur in \gls{rc}, \gls{rnist} and \gls{dr}. LFM is shared by \gls{rc} and \gls{dr}, and the six remaining phase codes by \gls{rnist} and \gls{dr}. \gls{rml} shares no class with any other dataset. 

For classes $a$ and $b$ to be equivalent, the model must be incapable of distinguishing them. In such a case, the model would reliably confuse them. We define the probability of misclassifying $a$ as $b$ as $\pi(a{\to}b)$  
\begin{equation}
\pi(a{\to}b) \;=\;
\frac{P(\hat{y}=b \mid y=a)}
     {\sum_{c \in \mathcal{S}_j} P(\hat{y}=c \mid y=a)},
\label{eq:share}
\end{equation}
where $\mathcal{S}_j$ is the class set of task $j$. $\pi$ is the fraction of $a$'s mass in task $j$ assigned to $b$. Normalising within task $j$ cancels both the fraction of $a$'s mass that leaves its own task and the overall attraction of task $j$. Chance is $1/|\mathcal{S}_j|$ and the empirical ceiling is $\pi(b{\to}b)$. From this, \eqref{eq:share} is corrected such that $\tilde\pi {\in} [-1,1]$ \eqref{eq:position}. $\tilde\pi=0$ is chance and $\tilde\pi=1$ is identity between $a$ and $b$. 
\begin{equation}
\tilde\pi(a \!\to\! b) \;=\;
\frac{\bar{\pi}(a \!\to\! b) - 1/|\mathcal{S}_j|}
     {\bar{\pi}(b \!\to\! b) - 1/|\mathcal{S}_j|},
\label{eq:position}
\end{equation}

\begin{table}[t]
\centering
\footnotesize
\setlength{\tabcolsep}{4pt}
\caption{Corrected probability of confusion $\tilde\pi$ \eqref{eq:position} of classes sharing a modulation. Values ${>}0$ indicate confusion rates are higher than chance, with a ceiling of $\tilde\pi{=}1$. ``---'' denotes that no mass is assigned to that task.}
\label{tab:overlap-pairs}
\resizebox{0.9\columnwidth}{!}{%
\begin{tabular}{llrrcrr}
\hline\toprule
& & \multicolumn{2}{c}{\textbf{JT-TIL}} & & \multicolumn{2}{c}{\textbf{JT-CIL}} \\
\cline{3-4}\cline{6-7}
\textbf{Class} & $\mathbf{a}$ \textbf{/} $\mathbf{b}$ & $\mathbf{a\!\to\!b}$ & $\mathbf{b\!\to\!a}$ & & $\mathbf{a\!\to\!b}$ & $\mathbf{b\!\to\!a}$ \\
\hline\toprule
P1          & RN 2 / DR 3 & $-$0.17 & $-$0.10 & & 0.67 & $-$0.11 \\
P3          & RN 1 / DR 2 & $-$0.04 & 0.30 & & 0.26 & $-$0.12 \\
Zadoff-Chu  & RN 2 / DR 2 & 0.03 & $-$0.03 & & 0.03 & 0.06 \\
Frank       & RC 1 / RN 1 & 0.12 & 0.00 & & $-$0.21 & $-$0.15 \\
P2          & RN 2 / DR 2 & $-$0.21 & 0.24 & & $-$0.24 & --- \\
Barker      & RC 1 / DR 1 & $-$0.21 & 0.00 & & $-$0.25 & --- \\
P4          & RN 2 / DR 4 & $-$0.28 & 0.09 & & $-$0.32 & --- \\
Barker      & RN 2 / DR 1 & $-$0.21 & $-$0.19 & & $-$0.25 & --- \\
Barker      & RC 1 / RN 2 & $-$0.18 & $-$0.23 & & $-$0.22 & $-$0.26 \\
Frank       & RN 1 / DR 3 & $-$0.22 & $-$0.16 & & $-$0.34 & $-$0.21 \\
LFM         & RC 1 / DR 2 & $-$0.16 & $-$0.25 & & $-$0.29 & --- \\
Px          & RN 1 / DR 1 & $-$0.22 & $-$0.19 & & $-$0.35 & $-$0.21 \\
Frank       & RC 1 / DR 3 & $-$0.27 & $-$0.25 & & $-$0.34 & $-$0.27 \\
\bottomrule\hline
\end{tabular}%
}
\end{table}

No pair is consistently confused in both directions. P1 shows a partial, unidirectional confusion wherein P1 of RN is confused with P1 of DR at $\tilde\pi{=}0.67$. However, this is only true for one direction and under CIL; the same result cannot be seen under TIL. All other pairs lie at or near chance with $\tilde\pi{\le}0.30$. This empirically shows that classes sharing a modulation across datasets are distinguishable. 

%% file: 07_Appendix/tid_leakage.tex
\subsection{Task Identity Leakage}
\label{apx:tid-leakage}

This appendix investigates whether a model indirectly learns \gls{tid}. The models tested are trained using \gls{tid} but evaluated across all task heads. Let $M[i,j]$ denote the fraction of task $i$'s predictions that fall in task $j$. The equivalent of \eqref{eq:position} at task level is
\begin{equation}
\tilde\pi(i \!\to\! j) \;=\; \frac{M[i,j] - c_{ij}}{1 - c_{ij}},
\qquad
c_{ij} \;=\; \operatorname*{mean}_{k \,\notin\, \{i,j\}} M[k,j],
\label{eq:block-position}
\end{equation}
where $c_{ij}$ is the fraction of task $j$'s predictions drawn from unrelated tasks. For TID retention ($j{=}i$), $\tilde\pi{=}0$ indicates chance, and $\tilde\pi{=}1$ indicates perfect task identification. Own-task $\tilde\pi$ exceeds zero for all tasks for both JT and Ring-ER, with means of $0.37$ and $0.22$, respectively. Retention is highest for RC under JT ($0.60$) and RN~1 under Ring-ER ($0.44$). However, interpreting this as TID leakage is erroneous: computing Spearman rank correlation between task difficulty (Table~\ref{tab:class-splits}) and TID retention gives $\rho{=}0.84$ and $0.88$ for JT and Ring-ER, respectively. This shows a very strong correlation between task difficulty and TID retention. 

\begin{table}[t]
\centering
\small
\setlength{\tabcolsep}{3.5pt}
\setlength{\aboverulesep}{0pt}
\setlength{\belowrulesep}{0pt}
\caption{Quantified leakage of TID on a scale of $\tilde\pi{\in}[-1,1]$ according to \eqref{eq:block-position}, averaged across $6$ seeds.}
\label{tab:block-position}
\resizebox{\columnwidth}{!}{
\begin{tabular}{l|rcccccccccc|c}
\hline\toprule
\multirow{2}{*}{\textbf{Method}} & \textbf{Dataset:} & \textbf{RC} & \multicolumn{2}{c}{\textbf{RN}} & \multicolumn{3}{c}{\textbf{\gls{rml}}} & \multicolumn{4}{c|}{\textbf{\gls{dr}}} & \multirow{2}{*}{\textbf{Mean}} \\
\cmidrule(lr){3-3} \cmidrule(lr){4-5} \cmidrule(lr){6-8} \cmidrule(lr){9-12}
& \textbf{Task:} & \textbf{1} & \textbf{1} & \textbf{2} & \textbf{1} & \textbf{2} & \textbf{3} & \textbf{1} & \textbf{2} & \textbf{3} & \textbf{4} & \\
\hline\toprule
\textbf{JT}      & $\mathbf{\pi{:}}$ & 0.60 & 0.51 & 0.54 & 0.51 & 0.32 & 0.11 & 0.33 & 0.39 & 0.17 & 0.23 & \textbf{0.37} \\
\textbf{Ring-ER} & $\mathbf{\pi{:}}$ & 0.38 & 0.44 & 0.29 & 0.26 & 0.27 & 0.06 & 0.13 & 0.20 & 0.13 & 0.04 & \textbf{0.22} \\
\bottomrule\hline
\end{tabular}%
}
\end{table}

%% file: 07_Appendix/hpt.tex
\subsection{Hyperparameter Selection} \label{sec:appendixA}
All algorithms were tuned via grid search over the parameter ranges listed in Table~\ref{tab:hpt}. The learning rate search space is common to all methods; algorithm-specific regularisation coefficients and auxiliary parameters were swept independently over the shown ranges. Separate hyperparameter selections were made for the \gls{til} and \gls{cil} settings, as the optimal configuration was found to differ non-trivially across the two protocols. Methods marked N/A are incompatible with \gls{cil}. Final selected values are reported alongside their respective sweep ranges.
\begin{table}
\centering
\caption{Hyperparameter sweep ranges for all \gls{cl} approaches.
A single \gls{lr}, selected on the iid upper bound, is shared by all approaches; ``fixed'' denotes a parameter pinned rather than swept.}
\label{tab:hpt}
\footnotesize
\setlength{\tabcolsep}{3pt}
\resizebox{0.85\columnwidth}{!}{%
\begin{tabular}{c||c|c|
                >{\centering\arraybackslash}p{1.2cm}|
                >{\centering\arraybackslash}p{1.2cm}}
\hline\toprule
\multirow{2}{*}{\textbf{Method}} & \multirow{2}{*}{\textbf{Parameter}} & \multirow{2}{*}{\textbf{Sweep Range}} & \multicolumn{2}{c}{\textbf{Selected Value}} \\
         &           &             & \textbf{TIL}         & \textbf{CIL} \\
\hline\toprule
All & LR & $\{1e^{-4},3e^{-4},\dots,1e^{-1},3e^{-1}\}$ & $3e^{-2}$ & $3e^{-2}$\\
\midrule
\multirow{2}{*}{BCL-Dual}       & $\lambda$     & $\{1e^{-1},5e^{-1},1,5,\dots,5e^{2}\}$
                                                & $1$ & $5e^{-1}$ \\
                                & $\beta$       & $\{1e^{-1},2e^{-1},\dots,1\}$
                                                & $2e^{-1}$ & $2e^{-1}$ \\
\midrule
\multirow{2}{*}{CTN}            & $\beta$       & $\{1e^{-4},3e^{-4},\dots,1e^{-1},3e^{-1}\}$
                                                & $1e^{-4}$ & \multirow{2}{*}{N/A} \\
                                & $\lambda$     & $\{1e^{-1},5e^{-1},1,5,\dots,5e^{2}\}$
                                                & $5$ &  \\
\midrule
\multirow{1}{*}{EWC}            & $\lambda$     & $\{1,1e^{1},\dots,1e^{9}\}$
                                                & $1e^{7}$ & $1e^{6}$ \\
\midrule
\multirow{1}{*}{GEM}            & $\gamma$     & $\{0,1e^{-1},2e^{-1},\dots,1\}$
                                                & $3e^{-1}$ & $1e^{-1}$ \\
\midrule
\multirow{2}{*}{HAT}            & $s_{max}$     & $\{5,1e^{1},1.5e^{1},2.5e^{1},5e^{1},\dots,8e^{2}\}$
                                                & $4e^{2}$ & \multirow{2}{*}{N/A} \\
                                & $\gamma$      & $\{1e^{-1},2.5e^{-1},1,1.5,2.5\}$
                                                & $1$ &  \\
\midrule
\multirow{2}{*}{La-MAML}        & $\alpha_0$    & N/A (LR equivalent) 
                                                & $3e^{-2}$ & $3e^{-2}$ \\
                                & $\eta$        & $\{1e^{-4},3e^{-4},\dots,1e^{-1},3e^{-1}\}$
                                                & $3e^{-2}$ & $1e^{-3}$ \\
\midrule
\multirow{2}{*}{C-MAML}         & $\beta$       & N/A (LR equivalent) 
                                                & $3e^{-2}$ & $3e^{-2}$ \\
                                & $\alpha_0$    & $\{1e^{-4},3e^{-4},\dots,1e^{-1},3e^{-1}\}$
                                                & $3e^{-3}$ & $1e^{-4}$ \\
                                
\midrule
\multirow{3}{*}{S-MAML}         & $\beta$       & N/A (LR equivalent) 
                                                & $3e^{-2}$ & $3e^{-2}$ \\
                                & $\alpha_0$    & $\{1e^{-4},3e^{-4},\dots,1e^{-1},3e^{-1}\}$
                                                & $1e^{-2}$ & $1e^{-2}$ \\
                                & $\eta$        & $\{1e^{-4},3e^{-4},\dots,1e^{-1},3e^{-1}\}$
                                                & $3e^{-4}$ & $1e^{-4}$ \\
                                
\midrule
\multirow{2}{*}{La-ER}          & $\eta$        & $\{1e^{-4},3e^{-4},\dots,1e^{-1},3e^{-1}\}$
                                                & $3e^{-2}$ & $3e^{-1}$ \\
                                & $\lambda$     & $\{1e^{-1},5e^{-1},1,5,\dots,5e^{2}\}$
                                                & $1$ & $1$ \\
\midrule
\multirow{1}{*}{LwF}            & $\lambda$     & $\{1e^{-1},5e^{-1},1,5,\dots,5e^{2}\}$
                                                & $5e^{-1}$ & $1e^{-1}$ \\
\midrule
\multirow{1}{*}{PackNet}        & $\rho$        & $\{5e^{-1},6e^{-1},\dots,9e^{-1}\}$
                                                & $8e^{-1}$ & N/A \\
\midrule
\multirow{2}{*}{RWalk}          & $\lambda$     & $\{1,1e^{1},\dots,1e^{6}\}$
                                                & $1e^{4}$ & $1e^{2}$ \\
                                & $\alpha$      & $\{1e^{-1},3e^{-1},\dots,9e^{-1}\}$
                                                & $7e^{-1}$ & $3e^{-1}$ \\
\midrule
\multirow{1}{*}{SI}             & $\lambda$     & $\{5e^{-1},1,1e^{1},\dots,1e^{6}\}$
                                                & $1e^{4}$ & $5e^{-1}$ \\
\midrule
\multirow{3}{*}{UCL}            & $r$           & $\{1e^{-3},2.5e^{-3},5e^{-3},\dots,5e^{-1}\}$
                                                & $2.5e^{-3}$ & $1e^{-1}$ \\
                                & $\beta$       & $\{1e^{-6},1e^{-5},\dots,1e^{1}\}$
                                                & $1e^{-6}$ & $1e^{-5}$ \\
                                & $\text{LR}_\rho$    & $\{1e^{-4},3e^{-4},\dots,1e^{-1}\}$
                                                & $2e^{-2}$ & $1e^{-3}$ \\
\bottomrule\hline
\end{tabular}%
}
\end{table}

%% file: 07_Appendix/full_ablations.tex
\subsection{Full Ablation Results}
\label{apx:full-ablations}
Tables~\ref{tab:til_ablations_1e_5e} and~\ref{tab:cil_ablations_1e_5e} report the complete ablation results summarised in Tables~\ref{tab:til_ablations} and~\ref{tab:cil_ablations}, including absolute $\fcl$, 95\% \glspl{ci} and Holm-corrected $p$-values.

\input{06_Ablation/Tables/til_ablations_1e_5e.tex}
\input{06_Ablation/Tables/cil_ablations_1e_5e.tex}

%% file: 06_Ablation/Tables/til_ablations_1e_5e.tex
\begin{table*}[t]
    \centering
    \footnotesize
    \caption{Full \gls{til} ablation results, summarised in Table~\ref{tab:til_ablations}. $\fcl$ is mean\,$\pm$\,sample std; $\Delta$F1 is the paired change against the family baseline with 95\% $t$ \gls{ci}; $p_{\mathrm{Holm}}$ is corrected within family. $^\dagger n{=}12$ seeds. $^\ddagger$Paired against A1.}
    \label{tab:til_ablations_1e_5e}
    \resizebox{\textwidth}{!}{
    \begin{tabular}{llcccccccccc}
    \hline\toprule
     & & \multicolumn{4}{c}{\textbf{Single-epoch}} & & \multicolumn{4}{c}{\textbf{Multi-epoch}} \\
    \cmidrule(lr){3-6} \cmidrule(lr){8-11}
    \textbf{ID} & \textbf{Mechanism altered} & $\mathbf{\fcl}$ & $\mathbf{\Delta}$\textbf{F1 [95\% CI]} & $\mathbf{p_{\mathrm{Holm}}}$ & \textbf{Verdict} & & $\mathbf{\fcl}$ & $\mathbf{\Delta}$\textbf{F1 [95\% CI]} & $\mathbf{p_{\mathrm{Holm}}}$ & \textbf{Verdict} \\
    \hline\toprule
    \multicolumn{11}{l}{\textit{GEM}} \\
    G0 & None (full method) & \pmv{62.83}{1.10} & -- & -- & -- &  & \pmv{68.54}{0.83} & -- & -- & -- \\
    G1 & Gradient projection $\rightarrow$ episodic replay (ring) & \pmv{58.07}{1.37} & $-4.76$ [$-5.63$, $-3.90$] & $<$0.001 & \textbf{SS} &  & \pmv{65.39}{1.01} & $-3.16$ [$-3.68$, $-2.63$] & $<$0.001 & \textbf{SS} \\
    \midrule
    \multicolumn{11}{l}{\textit{Res-ER}} \\
    E0 & None (full method) & \pmv{58.89}{1.91}$^\dagger$ & -- & -- & -- &  & \pmv{66.94}{0.70} & -- & -- & -- \\
    E1 & Reservoir $\rightarrow$ static ring buffer & \pmv{57.44}{1.68}$^\dagger$ & $-1.45$ [$-2.26$, $-0.64$] & 0.005 & \textbf{SS} &  & \pmv{65.39}{1.01} & $-1.56$ [$-2.23$, $-0.88$] & 0.004 & \textbf{SS} \\
    E2 & Reservoir $\rightarrow$ fully-utilised ring & \pmv{58.88}{1.39} & $-0.08$ [$-0.92$, $+0.77$] & 0.828 & NS &  & \pmv{66.40}{0.67} & $-0.54$ [$-0.96$, $-0.12$] & 0.022 & \textbf{SS} \\
    \midrule
    \multicolumn{11}{l}{\textit{C-MAML}} \\
    M0 & None (full method, first-order) & \pmv{59.71}{1.01}$^\dagger$ & -- & -- & -- &  & \pmv{63.91}{1.31}$^\dagger$ & -- & -- & -- \\
    M1 & Meta-batch averaging (3 $\rightarrow$ 1) & \pmv{60.56}{0.50} & $+0.62$ [$+0.07$, $+1.16$] & 0.068 & MS &  & \pmv{64.66}{1.20}$^\dagger$ & $+0.75$ [$+0.31$, $+1.19$] & 0.005 & \textbf{SS} \\
    M2 & Inner-loop adaptation ($\alpha$ $\rightarrow$ 0) & \pmv{59.65}{1.06}$^\dagger$ & $-0.06$ [$-0.59$, $+0.47$] & 0.802 & NS &  & \pmv{65.63}{0.86} & $+1.38$ [$+0.80$, $+1.97$] & 0.005 & \textbf{SS} \\
    \midrule
    \multicolumn{11}{l}{\textit{BCL-Dual}} \\
    B0 & None (full method) & \pmv{56.88}{1.74}$^\dagger$ & -- & -- & -- &  & \pmv{65.37}{1.40}$^\dagger$ & -- & -- & -- \\
    B1 & KL distillation & \pmv{55.14}{1.45} & $-1.52$ [$-2.36$, $-0.69$] & 0.027 & \textbf{SS} &  & \pmv{62.73}{1.19} & $-2.96$ [$-3.65$, $-2.28$] & $<$0.001 & \textbf{SS} \\
    B2 & Bilevel interpolation ($\beta$ $\rightarrow$ 1) & \pmv{55.26}{2.01} & $-1.40$ [$-2.70$, $-0.10$] & 0.078 & MS &  & \pmv{63.35}{3.01}$^\dagger$ & $-2.03$ [$-3.43$, $-0.62$] & 0.027 & \textbf{SS} \\
    B3 & Bilevel loop (matched budget) & \pmv{55.37}{1.60}$^\dagger$ & $-1.51$ [$-3.02$, $-0.01$] & 0.078 & MS &  & \pmv{65.59}{2.08}$^\dagger$ & $+0.22$ [$-0.54$, $+0.97$] & 1.000 & NS \\
    B4 & Validation buffer $\rightarrow$ current batch (outer step) & \pmv{54.77}{2.65}$^\dagger$ & $-2.11$ [$-3.60$, $-0.63$] & 0.039 & \textbf{SS} &  & \pmv{65.23}{2.51}$^\dagger$ & $-0.14$ [$-1.38$, $+1.10$] & 1.000 & INC \\
    \midrule
    \multicolumn{11}{l}{\textit{CTN}} \\
    T0 & None (full method) & \pmv{57.17}{1.03}$^\dagger$ & -- & -- & -- &  & \pmv{61.63}{1.03}$^\dagger$ & -- & -- & -- \\
    T1 & FiLM task conditioning & \pmv{56.55}{1.15}$^\dagger$ & $-0.62$ [$-0.95$, $-0.28$] & 0.004 & \textbf{SS} &  & \pmv{61.35}{1.40}$^\dagger$ & $-0.27$ [$-0.86$, $+0.31$] & 0.322 & NS \\
    T2 & KL distillation & \pmv{56.15}{0.97} & $-1.17$ [$-1.68$, $-0.65$] & 0.004 & \textbf{SS} &  & \pmv{59.21}{1.40} & $-2.88$ [$-3.87$, $-1.88$] & 0.001 & \textbf{SS} \\
    \midrule[\heavyrulewidth]
    \multicolumn{11}{l}{\textit{GEM$^+$}} \\
    A0 & Baseline method (GEM) & \pmv{62.83}{1.10} & -- & -- & -- &  & \pmv{68.54}{0.83} & -- & -- & -- \\
    A1 & \textit{(added)} Fully-utilised (dynamic) ring & \pmv{63.95}{1.73} & $+1.12$ [$+0.15$, $+2.10$] & 0.032 & \textbf{SS} &  & \pmv{69.19}{0.66} & $+0.65$ [$+0.21$, $+1.09$] & 0.012 & \textbf{SS} \\
    A2 & \textit{(added)} KL distillation & \pmv{65.60}{1.64} & $+2.77$ [$+1.43$, $+4.11$] & 0.006 & \textbf{SS} &  & \pmv{72.75}{0.84} & $+4.21$ [$+3.71$, $+4.70$] & $<$0.001 & \textbf{SS} \\
    C1$^\ddagger$ & \textit{(added)} Dynamic ring + KL distillation & \pmv{66.51}{2.29} & $+2.56$ [$+1.68$, $+3.44$] & 0.002 & \textbf{SS} &  & \pmv{73.28}{0.94} & $+4.09$ [$+3.67$, $+4.51$] & $<$0.001 & \textbf{SS} \\
    C1$^\star$ & \textit{(added)} Dynamic ring + KL distillation, held-out seeds (GEM$^+$) & \pmv{66.17}{1.65} & $+3.39$ [$+3.03$, $+3.74$] & $<$0.001 & \textbf{SS} &  & \pmv{74.22}{0.95} & $+5.52$ [$+4.93$, $+6.11$] & $<$0.001 & \textbf{SS} \\
    \midrule
    \multicolumn{11}{l}{\textit{Res-ER$^+$}} \\
    E0 & Baseline method (Res-ER) & \pmv{58.89}{1.91}$^\dagger$ & -- & -- & -- &  & \pmv{66.94}{0.70} & -- & -- & -- \\
    R1 & \textit{(added)} KL distillation & \pmv{62.10}{2.86} & $+3.15$ [$+1.83$, $+4.47$] & 0.005 & \textbf{SS} &  & \pmv{71.69}{1.41} & $+4.75$ [$+3.50$, $+6.00$] & $<$0.001 & \textbf{SS} \\
    R1$^\star$ & \textit{(added)} KL distillation, held-out seeds (Res-ER$^+$) & \pmv{61.70}{1.88} & $+2.55$ [$+1.08$, $+4.02$] & 0.007 & \textbf{SS} &  & \pmv{71.91}{1.33} & $+4.61$ [$+3.91$, $+5.31$] & $<$0.001 & \textbf{SS} \\
    \bottomrule\hline
    \end{tabular}}
    \end{table*}

%% file: 06_Ablation/Tables/cil_ablations_1e_5e.tex
\begin{table}[t]
    \centering
    \footnotesize
    \caption{Full \gls{cil} ablation results, summarised in Table~\ref{tab:cil_ablations}. $\fcl$ is mean\,$\pm$\,sample std; $\Delta$F1 is the paired change against the family baseline with 95\% $t$ \gls{ci}; $p_{\mathrm{Holm}}$ is corrected within family. $^\dagger n{=}12$ seeds.}
    \label{tab:cil_ablations_1e_5e}
    \resizebox{1\columnwidth}{!}{%
    \begin{tabular}{lcccccccccc}
    \hline\toprule
    & \multicolumn{4}{c}{\textbf{Single-epoch}} & & \multicolumn{4}{c}{\textbf{Multi-epoch}} \\
    \cmidrule(lr){2-5} \cmidrule(lr){7-10}
    \textbf{ID} & $\mathbf{\fcl}$ & $\mathbf{\Delta}$\textbf{F1 [95\% CI]} & $\mathbf{p_{\mathrm{Holm}}}$ & \textbf{Verdict} & & $\mathbf{\fcl}$ & $\mathbf{\Delta}$\textbf{F1 [95\% CI]} & $\mathbf{p_{\mathrm{Holm}}}$ & \textbf{Verdict} \\
    \hline\toprule
    \multicolumn{10}{l}{\textit{Res-ER}} \\
    E0 & \pmv{31.13}{6.11}$^\dagger$ & -- & -- & -- &  & \pmv{40.50}{1.65}$^\dagger$ & -- & -- & -- \\
    E1 & \pmv{32.08}{2.35}$^\dagger$ & $+0.95$ [$-1.79$, $+3.69$] & 0.923 & INC &  & \pmv{36.37}{1.63} & $-4.45$ [$-5.69$, $-3.21$] & $<$0.001 & \textbf{SS} \\
    E2 & \pmv{31.94}{3.52}$^\dagger$ & $+0.81$ [$-1.66$, $+3.27$] & 0.923 & INC &  & \pmv{38.47}{1.80} & $-2.36$ [$-3.19$, $-1.52$] & 0.002 & \textbf{SS} \\
    \midrule
    \multicolumn{10}{l}{\textit{C-MAML}} \\
    M0 & \pmv{37.17}{4.30}$^\dagger$ & -- & -- & -- &  & \pmv{41.86}{1.84}$^\dagger$ & -- & -- & -- \\
    M1 & \pmv{35.06}{2.95} & $-3.72$ [$-7.38$, $-0.07$] & 0.095 & MS &  & \pmv{41.38}{1.70}$^\dagger$ & $-0.48$ [$-1.32$, $+0.35$] & 0.459 & INC \\
    M2 & \pmv{32.53}{3.22} & $-6.25$ [$-10.03$, $-2.48$] & 0.024 & \textbf{SS} &  & \pmv{40.65}{1.73}$^\dagger$ & $-1.22$ [$-1.89$, $-0.54$] & 0.006 & \textbf{SS} \\
    \midrule
    \multicolumn{10}{l}{\textit{BCL-Dual}} \\
    B0 & \pmv{37.47}{1.44}$^\dagger$ & -- & -- & -- &  & \pmv{42.26}{1.69}$^\dagger$ & -- & -- & -- \\
    B1 & \pmv{36.43}{1.47} & $-1.88$ [$-3.39$, $-0.38$] & 0.047 & MS &  & \pmv{39.92}{1.21} & $-2.54$ [$-3.22$, $-1.85$] & 0.001 & \textbf{SS} \\
    B2 & \pmv{37.42}{3.00}$^\dagger$ & $-0.05$ [$-1.57$, $+1.47$] & 0.944 & INC &  & \pmv{42.77}{1.47}$^\dagger$ & $+0.51$ [$-0.17$, $+1.19$] & 0.128 & INC \\
    B3 & \pmv{31.50}{2.04} & $-6.81$ [$-8.81$, $-4.80$] & 0.002 & \textbf{SS} &  & \pmv{39.58}{2.00} & $-2.87$ [$-5.02$, $-0.73$] & 0.037 & \textbf{SS} \\
    B4 & \pmv{40.61}{0.61} & $+2.30$ [$+0.91$, $+3.70$] & 0.025 & \textbf{SS} &  & \pmv{46.30}{0.49} & $+3.84$ [$+1.96$, $+5.72$] & 0.011 & \textbf{SS} \\
    \midrule[\heavyrulewidth]
    \multicolumn{10}{l}{\textit{C-MAML$^+$ (C-MAML + KL distillation)}} \\
    M0 & \pmv{37.17}{4.30}$^\dagger$ & -- & -- & -- &  & \pmv{41.86}{1.84}$^\dagger$ & -- & -- & -- \\
    M0$^\S$ & \pmv{38.78}{1.91} & -- & -- & -- &  & \pmv{41.58}{2.26} & -- & -- & -- \\
    C1 & \pmv{41.39}{2.19} & $+2.60$ [$+1.39$, $+3.81$] & 0.008 & \textbf{SS} &  & \pmv{45.18}{2.50} & $+3.61$ [$+2.07$, $+5.14$] & 0.002 & \textbf{SS} \\
    C1$^\star$ (C-MAML$^+$) & \pmv{41.09}{0.85} & $+2.72$ [$+1.22$, $+4.22$] & 0.006 & \textbf{SS} &  & \pmv{45.04}{1.74} & $+2.67$ [$+1.49$, $+3.84$] & 0.002 & \textbf{SS} \\
    \midrule
    \multicolumn{10}{l}{\textit{BCL$^+$ (BCL-Dual + global reservoir + B4)}} \\
    B0 & \pmv{37.47}{1.44}$^\dagger$ & -- & -- & -- &  & \pmv{42.26}{1.69}$^\dagger$ & -- & -- & -- \\
    B4$^\star$ & \pmv{40.35}{1.01} & $+2.27$ [$+0.69$, $+3.84$] & 0.014 & MS &  & \pmv{46.08}{1.44} & $+2.74$ [$+1.40$, $+4.09$] & 0.003 & \textbf{SS} \\
    B5 & \pmv{39.82}{1.31} & $+1.51$ [$+0.80$, $+2.22$] & 0.003 & \textbf{SS} &  & \pmv{44.41}{1.46} & $+1.96$ [$+1.01$, $+2.90$] & 0.003 & \textbf{SS} \\
    B6 & \pmv{42.64}{1.48} & $+4.33$ [$+2.63$, $+6.03$] & 0.002 & \textbf{SS} &  & \pmv{48.79}{1.86} & $+6.34$ [$+4.24$, $+8.43$] & 0.001 & \textbf{SS} \\
    B6$^\star$ (BCL$^+$) & \pmv{41.12}{1.37} & $+3.04$ [$+1.00$, $+5.08$] & 0.012 & \textbf{SS} &  & \pmv{48.23}{1.33} & $+4.89$ [$+3.33$, $+6.44$] & $<$0.001 & \textbf{SS} \\
    \bottomrule\hline
    \end{tabular}}
\end{table}